\documentclass[10pt,twocolumn]{article}
\usepackage{geometry}
\usepackage{titlesec}
\titleformat{\section}{\normalfont\fontsize{10}{12}\bfseries}{\thesection}{1em}{}
\titleformat{\subsection}{\normalfont\fontsize{10}{12}\bfseries}{\thesubsection}{1em}{}
\usepackage[utf8]{inputenc}
\usepackage{graphicx}% Include figure files
\usepackage{dcolumn}% Align table columns on decimal point
\usepackage{amsmath}
\usepackage[figurename=Fig.,labelfont=bf,labelsep=space,font=small]{caption}
\usepackage[dvipsnames]{xcolor}
\usepackage[colorlinks=true, linkcolor=blue, urlcolor=blue]{hyperref}
\usepackage[numbers,sort&compress]{natbib}
\makeatletter
\def\NAT@bibstyle@naturemag{\NAT@num}
\makeatother
\usepackage{booktabs}
\usepackage{authblk}
\usepackage{amsfonts}
\usepackage{color,soul}
\renewcommand{\thesection}{\arabic{section}}

\renewcommand{\thesubsection}{\arabic{section}.\arabic{subsection}}
\renewcommand{\thefigure}{\arabic{figure}}
\renewcommand{\thetable}{\arabic{table}}
\hypersetup{%
  colorlinks,
  plainpages=false,
  breaklinks,
  pdfview=Fit,
  bookmarksopen,
  bookmarksnumbered,
  linkcolor=blue,
  anchorcolor=black,
  citecolor=blue,
  filecolor=black,
  }
\usepackage{orcidlink}
\newcommand{\orcid}[1]{\href{https://orcid.org/#1}{\orcidlink{#1}}}
\providecommand{\degree}{\ensuremath{^\circ}}
\usepackage{adjustbox}

\title{\textbf{High-performance silicon--metal laser welding resisting extreme conditions}}

\author[1]{Jiawei~Yin \orcid{0009-0007-4494-0394}}
\author[1]{Markus~Blothe \orcid{0000-0002-0890-2186}}
\author[1]{Hagen~P.~Kohl \orcid{0000-0002-4492-4454}}
\author[2]{Christina~Schütze \orcid{0009-0004-8948-1919}}
\author[1,3]{Stefan~Nolte \orcid{0000-0002-2919-2662}}
\author[1]{Maxime~Chambonneau \orcid{0000-0003-1910-2296} *}

\affil[1]{Friedrich Schiller University Jena, Institute of Applied Physics, Abbe Center of Photonics, Albert-Einstein-Straße 15, 07745 Jena, Germany}
\affil[2]{IL Metronic Sensortechnik GmbH, Mittelstraße 33, 98693 Ilmenau-Unterpörlitz, Germany}
\affil[3]{Fraunhofer Institute for Applied Optics and Precision Engineering IOF, Center of Excellence in Photonics, Albert-Einstein-Straße 7, 07745 Jena, Germany}
\affil[*]{\href{mailto:maxime.chambonneau@uni-jena.de}{\textcolor{blue}{maxime.chambonneau@uni-jena.de}} and \href{mailto:maxime.chambonneau@hotmail.fr}{\textcolor{blue}{maxime.chambonneau@hotmail.fr}}}

\date{}

\graphicspath{{Figs/}}

\begin{document}

\twocolumn[\begin{@twocolumnfalse}
\maketitle

\begin{abstract}
Reliable material joining is essential for countless industrial applications. While femtosecond laser welding provides a route beyond conventional bonding methods, demonstrations of silicon--metal joints are rare due to nonlinear propagation effects and have so far been limited to shear joining strengths of a few MPa.
Here, we demonstrate high-strength silicon--Kovar laser welding using sub-nanosecond pulses.
By optimizing the focal position, the welding pattern, the laser polarization, and the metal roughness, remarkable shear joining strengths up to $15.8$~MPa are achieved.
The silicon--metal joints withstand temperatures of up to 500\,\degree C and are hermetically sealed.
Together with the remarkable strength values, the resistance to harsh environments underpins the applicability of silicon--metal welding in various fields including aerospace, nuclear science, and metallurgy.\\
\end{abstract}
\end{@twocolumnfalse}]

\section*{Introduction}
\label{sec:intro}

Since its inception in the 2000s, ultrafast laser welding has attracted considerable attention because it can join a wide variety of similar \cite{Tamaki2005} and dissimilar \cite{Ozeki2008} materials.
The clean, fast, and contactless aspects of this emerging technique make it a robust alternative to traditional bonding methods such as adhesive bonding, eutectic bonding, and soldering.
Ultrafast laser welding relies on propagation in a first workpiece, energy deposition at the interface with a second workpiece, in turn leading to a temperature rise until both materials melt, mix, and eventually create strong bonds when resolidifying.
It was shown to perform well with ultrashort laser pulses when the first workpiece consists of a wide-bandgap material such as glass, polymer, ceramic, or sapphire \cite{Mingareev2012,Carter2014,Zhang2015,Richter2015,Richter2016,Carter2017,Gstalter2017,Zhang2018,Penilla2019,Gstalter2019,Cvecek2019,Hecker2020a,Sahoo2020,Hecker2020b,Capodacqua2023,Dzipalski2024,Morawska2024,Huo2025,Jia2025,Jiang2025,Su2025,Li2025a,Li2025b,Wang2025b,Li2025,Ji2025,Yu2026}.
However, until recently, this technique was not applicable in the silicon--metal configuration \cite{Chambonneau2021a}.
This is due to delocalized and decreased interfacial energy deposition as a consequence of nonlinear propagation in silicon for excessively short laser pulses \cite{Zavedeev2016,Chanal2017,Mareev2020,Chambonneau2021b,Ganguly2024,Wang2025,Chambonneau2026,Xie2026}.
Solutions that have led to through-silicon welding include precompensating for the nonlinear focal shift \cite{Chambonneau2021a}, enhancing the interfacial absorption with a metallic nanolayer \cite{Chambonneau2023}, and employing continuous irradiation \cite{Sari2008} or nanosecond pulses \cite{Sopena2022,vanAbeelen2026}.
Despite all these efforts, in the silicon--metal laser-welding configuration, single-digit MPa shear joining strengths have been demonstrated \cite{Chambonneau2021a}, and their stability under extreme temperature conditions and vacuum environment has not been investigated so far.\\

In this article, we demonstrate silicon--metal laser welding with sub-nanosecond pulses, which exhibits shear joining strengths up to $15.8$~MPa, i.e., more than $7 \times$ the preceding record of 2.2~MPa \cite{Chambonneau2021a}.
This achievement relies on an optimization of the focal position, the welding pattern, the laser polarization, and the metal roughness.
The applicability of our technique to harsh-environment conditions is then studied.
The silicon--metal joints combine high mechanical strength with thermal resistance up to 500\,\degree C and hermetic sealing.
The high strength values together with the temperature resistance and hermeticity indicate that the proposed silicon--metal laser welding technique could be implemented in industrial sectors such as aerospace, where joints are exposed to extreme conditions and other joining techniques may underperform.

\section{Experimental methods}
\label{sec:expmethods}

\begin{figure}[!ht]
\centering
\includegraphics[width=\linewidth]{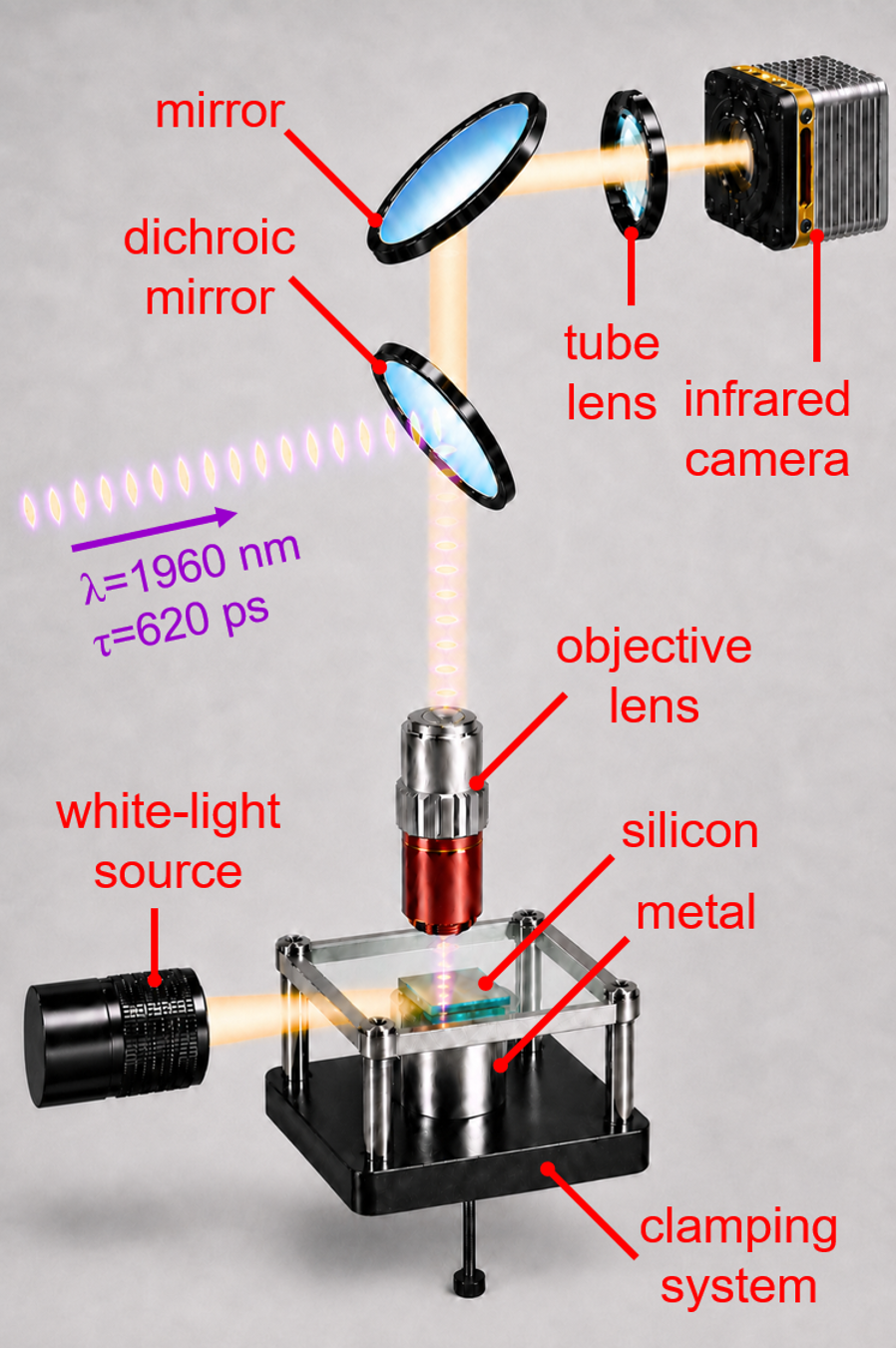}
\caption{\label{fig:Fig1} Schematic of the laser welding setup.}
\end{figure}

Laser welding experiments have been performed with the setup illustrated in Fig.~\ref{fig:Fig1}, which relies on a Tm:doped fiber laser delivering 620-ps pulses at a center wavelength of 1960~nm and a repetition rate of $\Omega=6.25$~kHz to avoid cumulative effects on a pulse-to-pulse basis.
The laser pulses exhibiting Gaussian temporal and spatial intensity profiles are focused with an objective lens of numerical aperture $\rm{NA}=0.26$ (Mitutoyo, M Plan Apo NIR $10\times$), leading to a focal spot diameter at $1/e^{2}$ of $2w_{0}=6.8$~$\mu$m, and a confocal parameter of $2z_{0}=170.9$~$\mu$m, as calculated in silicon with our vectorial model \textit{InFocus} \cite{Li2021,Li2021a}.
The input pulse energy $E_{\rm{in}}$, measured after the objective lens, is adjusted using a half-wave plate combined with a linear polarizer.\\

The beam is focused near the interface between a microelectronic-grade crystalline silicon sample [undoped, $5 \times 5 \times 1$~mm$^{3}$, (100)-oriented, double-side polished, total thickness variation $< 10$~$\mu$m, bow $< 30$~$\mu$m, and wrap $< 30$~$\mu$m], and a Kovar plate (12.5-mm thick cylinder with a diameter of 30~mm).
Kovar, i.e., an iron–nickel–cobalt alloy originally developed to exhibit a coefficient of thermal expansion similar to that of borosilicate glass, has been selected as a metal of interest because of its high melting point and low thermal expansivity, as discussed in Section~\ref{ssec:temperature}.
The optical contact between the two polished workpieces is guaranteed by cleaning the surface with isopropyl alcohol before laser processing, and maintaining a constant force by means of a clamping system throughout welding.
The clamping system described in Refs.~\cite{Li2025a,Li2025b} consists of a metal plate and a 6.5-mm thick optical glass plate (measured transmission of $90\%$ including linear absorption and Fresnel losses at both air--glass interfaces), whose separation can be adjusted to clamp the silicon--metal stack tightly between them.\\

Different welding patterns have been investigated.
To compare these patterns, a constant scanning speed of $v=2$~mm/s was selected, which corresponds to $N = 2w_{0}\Omega/v =21$ applied pulses per point of the pattern.
The samples and clamping system are positioned in the $xy$ plane perpendicular to the optical axis $z$ with a nanopositioning system (Aerotech, ANT130-160-XY).
The focusing objective lens is positioned along the $z$ axis with another nanopositioning stage (Aerotech, ANT130-060-L).
A customized dark-field infrared (IR) microscope working in reflection is employed for \textit{in-situ} monitoring of the welding.
This relies on broadband illumination emitted by a quartz tungsten-halogen lamp (Thorlabs, QTH10/M) at a grazing angle with respect to the samples.
The light scattered at the silicon--metal interface is collected by the same objective lens as the one used for focusing the laser pulses, then transmitted by a dichroic mirror, and finally directed to the IR detection unit consisting of a tube lens (Thorlabs, TTL200-S8) and an InGaAs camera (Xenics, Bobcat 320).
The latter represents a major improvement compared to our previous work where a silicon-based camera was used to position the geometrical focus with respect to the entrance surface of silicon \cite{Chambonneau2021a,Chambonneau2023}.
Indeed, the InGaAs-based sensor allows for direct through-silicon \textit{in-situ} observation of modifications at the interface (defined as $z=0$).\\

\begin{figure}[!ht]
\centering
\includegraphics[width=\linewidth]{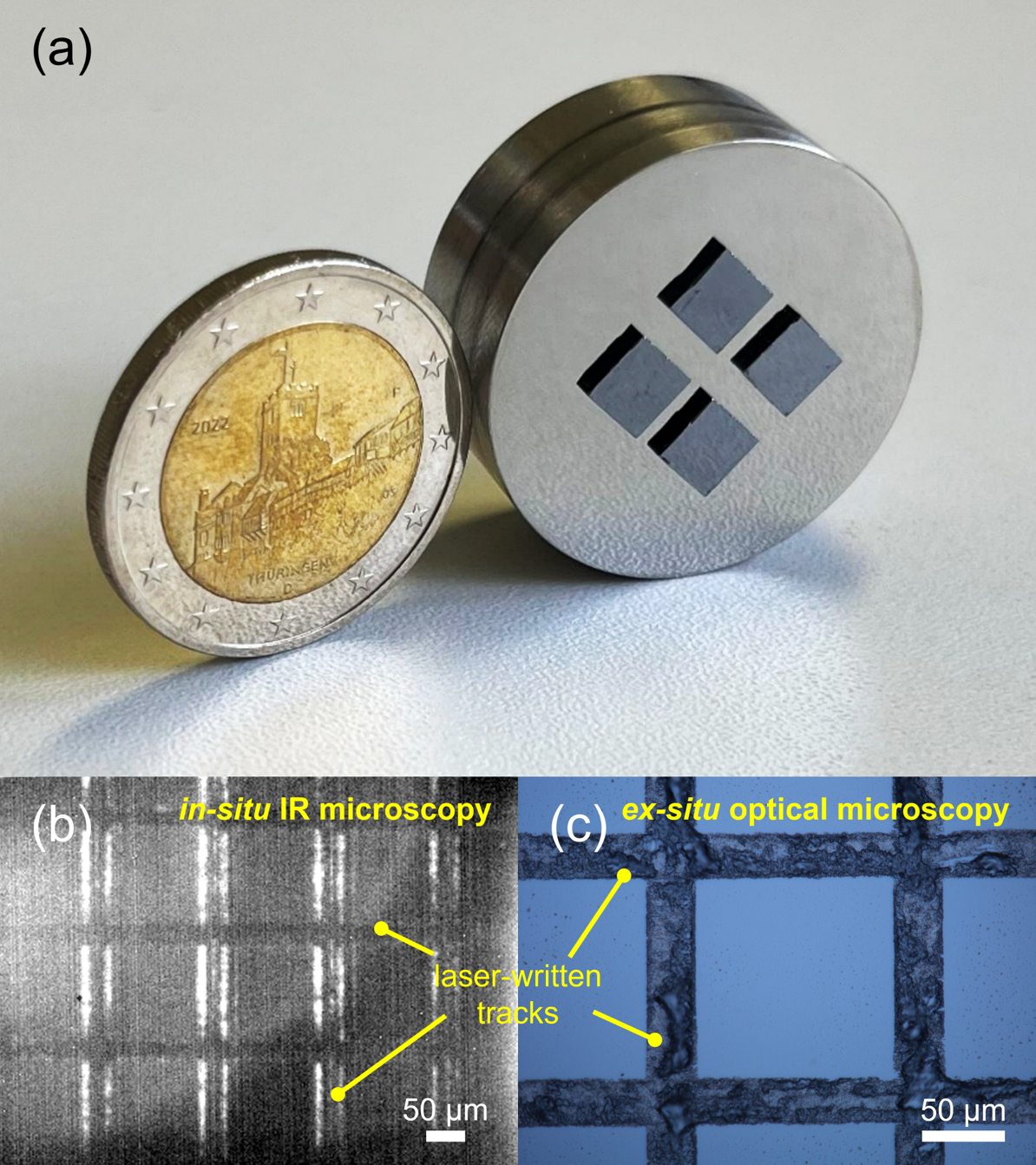}
\caption{\label{fig:Fig2} Typical silicon--metal welds. (a) Photograph of four $5 \times 5$~mm$^{2}$ silicon samples joined to Kovar. (b) \textit{In-situ} infrared (IR) microscopy, and (c) \textit{ex-situ} optical microscopy images of the interface. The observations in (b) and (c) are performed through silicon, and on the silicon sample after sample separation, respectively. Pulse energy: $E_{\rm{in}}=6$~$\mu$J, focal position: $z = 380$~$\mu$m, welding pattern: grid, laser polarization: linear, metal roughness: $R_{q}=2.8$~nm.}
\end{figure}

After welding, the samples are bonded together, as shown in Fig.~\ref{fig:Fig2}(a).
The welding performance is evaluated \textit{ex-situ} in terms of shear joining strength.
Breaking tests are performed by applying a force onto the edge of the silicon sample with an indenter until it is separated from the metal.
The corresponding peak breaking force is recorded by a force gauge (RS PRO, 111-3690).
The shear joining strength is obtained by calculating the ratio between the measured peak breaking force and the apparent welded area, corresponding to the laser-written tracks and determined with \textit{ex-situ} optical microscopy characterization, which are consistent with \textit{in-situ} observations [Fig.~\ref{fig:Fig2}(b) and (c)].
For each experimental condition involving shear-strength measurements, at least five independent silicon--metal joints were produced and mechanically tested.
The corresponding data points reported throughout the manuscript represent the average shear joining strength, while the error bars indicate the standard deviation over these independent measurements.\\

Prior to laser welding, the Kovar samples were polished to obtain different surface finishes using a polishing machine (Struers, LaboForce-100) with abrasives of different grain sizes.
A three-dimensional optical surface profiler (Zygo, NewView\texttrademark{} 9000) is employed after polishing to determine the root mean square (RMS) surface roughness $R_{q}$ over an $862 \times 862$-$\mu$m$^{2}$ area.
The typical metal roughness $R_{q}$ lies in the 2.8--466~nm range.\\

To evaluate the thermal resistance of the welds, high-temperature exposure is performed in a furnace [Linn Elektro Therm, FK-270/270/420/1300 (SO 1419)] for one hour at a set temperature $T$ in the 20--700\,\degree C range.
After high-temperature exposure, the samples cool down to room temperature for an additional hour before measuring the shear joining strength.\\

Finally, to determine the hermeticity of the joints, a 1-mm diameter hole was drilled through five metal samples.
One surface was then polished prior to the welding, so that the silicon sample fully covers one end of the hole.
The hermeticity of the welds was then evaluated with helium leak tests using a vacuum-box configuration with helium spraying on the opposite side of the joint, corresponding to principle B.2.2 of prEN 1779:2024 / DIN EN 1779:2024 draft \cite{standardleakrate}.
The leak rate was quantified using a helium mass-spectrometer leak detector (Leybold GmbH, PHOENIX 4.0 Quadro Wet).\\

\section{Results and Discussion}
\label{sec:resultsdiscussion}

\subsection{Optimizing the mechanical performance}
\label{ssec:optimization}

In this section, various laser and material parameters are optimized to obtain silicon--metal joints with the best performance in terms of shear joining strength.
These parameters include the focal position, the welding pattern, the laser polarization as well as the metal roughness.
All the experiments have been carried out for an input pulse energy of $E_{\rm{in}}=6$~$\mu$J, which corresponds to the maximum value our setup can deliver.
Two aspects motivated the choice of the highest possible pulse energy.
First, as shown in our previous study \cite{Chambonneau2021a}, the energy reaching the silicon--metal interface increases with $E_{\rm{in}}$, whereas the maximum fluence saturates due to nonlinear propagation effects.
Second, to ensure a homogeneous welding seam, the applied pulse energy has to be much higher than the interfacial modification threshold.
This aspect has been verified with modification tests (see Appendix~A).
The interfacial modification threshold is $675 \pm 25$~nJ, and a modification probability of $100\%$ is reached for $850 \pm 50$~nJ---both values being significantly below $E_{\rm{in}}=6$~$\mu$J.

\subsubsection{Focal position}
\label{sssec:energy_z}

In our previous work, we have shown that a precompensation for the nonlinear focal shift was required to optimize energy deposition at the silicon--metal interface with 10-ps pulses \cite{Chambonneau2021a}.
This is due to nonlinear propagation effects---in particular, the Kerr effect.
By employing much longer pulses (620~ps) in the present study, the laser intensity is decreased, and one could thus expect the propagation nonlinearities to be less severe.\\

\begin{figure}[!ht]
\centering
\includegraphics[width=\linewidth]{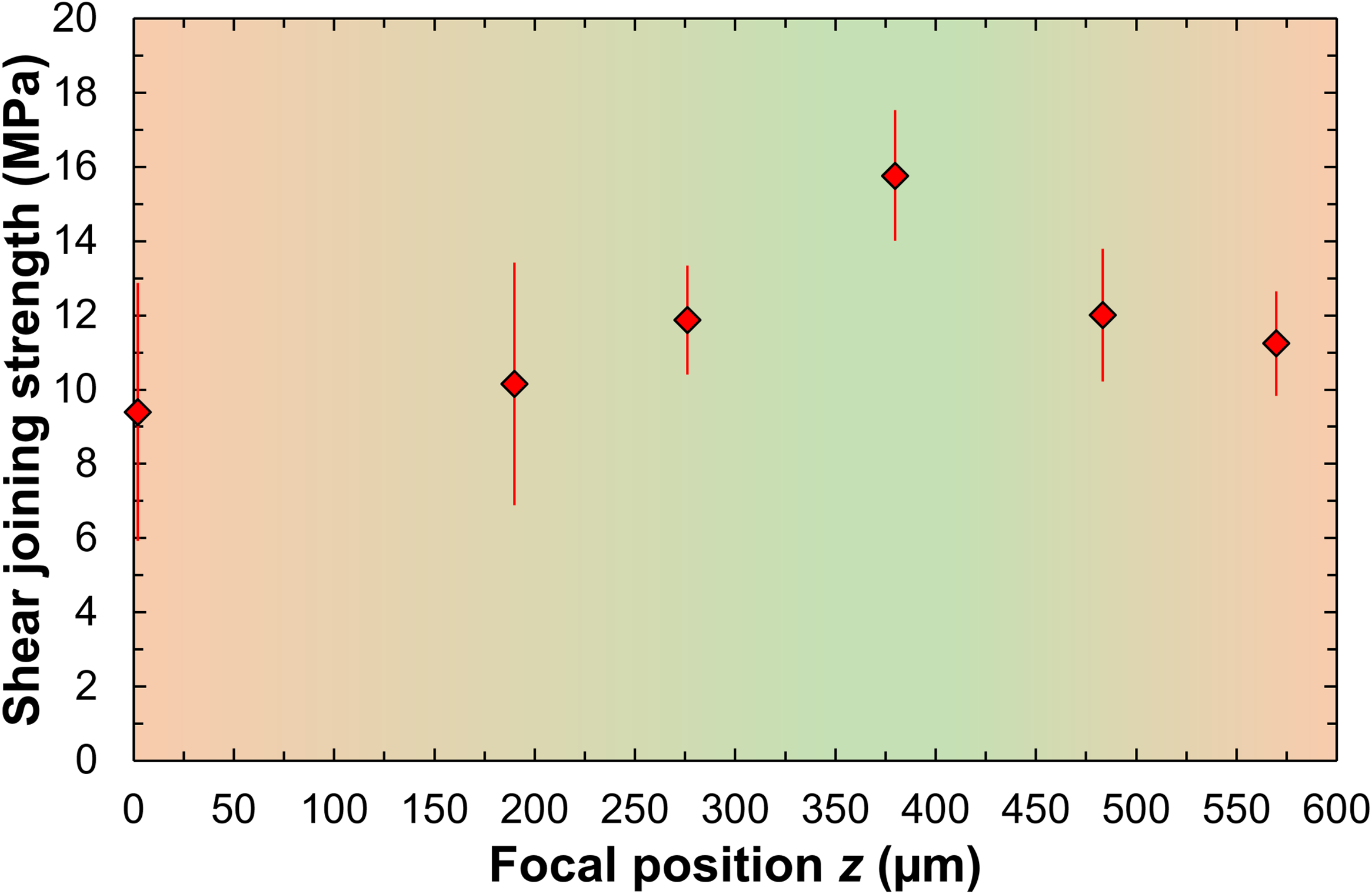}
\caption{\label{fig:Fig3} Evolution of the shear joining strength as a function of the focal position $z$ in silicon. The repeatability error bars indicate the standard deviation on the measured shear joining strength. The value $z=0$ corresponds to the silicon--metal interface, and $z > 0$ corresponds to a geometrical focus positioned in the metal. Pulse energy: $E_{\rm{in}}=6$~$\mu$J, welding pattern: grid, laser polarization: linear, metal roughness: $R_{q}=2.8$~nm.}
\end{figure}

To examine this, we have carried out silicon--metal welding experiments for various focal positions $z$ in silicon.
The silicon--metal interface is defined as $z=0$, and $z>0$ corresponds to a focus theoretically positioned after the silicon sample, i.e., in the metal.
The welding results are displayed in terms of shear joining strength in Fig.~\ref{fig:Fig3}.
When the geometrical focus is positioned at the interface ($z=0$), a shear joining strength of 9.4~MPa is already measured.
The strength value is further increased when the focus is moved downstream along the beam path ($0 < z \leq 380$~$\mu$m), reaching a peak value of 15.8~MPa for $z=380$~$\mu$m, before decreasing when the focus is too far from the interface ($z>380$~$\mu$m).
The optimal value of $z=380$~$\mu$m is consistent with previous studies which have shown that focal repositioning by several hundred micrometers can be required to achieve laser welding in different material configurations \cite{Carter2017,Morawska2024,Huo2025}.
However, it must be emphasized that the optimal $z$ value strongly depends on the laser parameters, such as beam size, pulse energy and pulse duration, the materials employed, and the mechanisms responsible for the focal shift.
From an engineering perspective, the need for a focal repositioning shown in Fig.~\ref{fig:Fig3} is an important result as it shows that the performance of silicon--metal welding can be improved by $68\%$ when the focal position is $z=380$~$\mu$m compared to the non-optimal configuration $z=0$.\\

Two mechanisms that can both lead to a focal shift may explain the existence of an optimal focusing depth.
First, accounting for Fresnel losses at the air--glass and glass--silicon interfaces, the peak power after the front surface of the silicon sample is $P_{\rm{peak}} \approx 7.4$~kW.
Based on nonlinear refractive index measurements ($n_{2}^{\rm{Si}} \approx 1.05 \times 10^{-17}$~m$^{2}$/W \cite{Bristow2007}), this power value is lower than the critical power for self-focusing in silicon $P_{\rm{cr}} \approx 15.9$~kW.
However, this estimate only accounts for the instantaneous Kerr response, and neglects the delayed medium response, which was shown to be non-negligible for picosecond pulses \cite{Chambonneau2026}.
Therefore, the critical power for self-focusing could be lower than the above-calculated value, and the Kerr effect may cause the nonlinear focal shift analogously to Refs.~\cite{Chambonneau2021a,Chambonneau2023}.
A second mechanism that can cause the nonlinear focal shift is thermal lensing.
Indeed, in contrast to studies using 10-ps laser pulses \cite{Chambonneau2021a,Chambonneau2023}, the pulse duration of 620~ps in this work implies that heat is substantially generated during the pulse \cite{Sundaram2002,Gattass2008}.
The resulting heating raises the refractive index, turning the heated zone into a time-dependent convex lens.\\

\subsubsection{Welding pattern and laser polarization}
\label{sssec:pattern}

Generally speaking, the welding pattern affects how residual stress builds up and is distributed spatially and temporally between the two workpieces \cite{Richter2016}.
Early works suggested that an Archimedean spiral pattern reduces stress accumulation \cite{Carter2014}; this pattern was therefore applied to different material configurations \cite{Carter2017,Li2025a,Li2025b}.
Nevertheless, a recent study suggested that the spiral pattern was not the most appropriate pattern for optimizing the shear joining strength, and that stronger bonds were created in glass-ceramics with a grid pattern \cite{Yu2026}.
Therefore, in our study which aims at optimizing silicon--metal welding, it is necessary to examine which pattern leads to the strongest welds.\\

\begin{figure}[!ht]
\centering
\includegraphics[width=\linewidth]{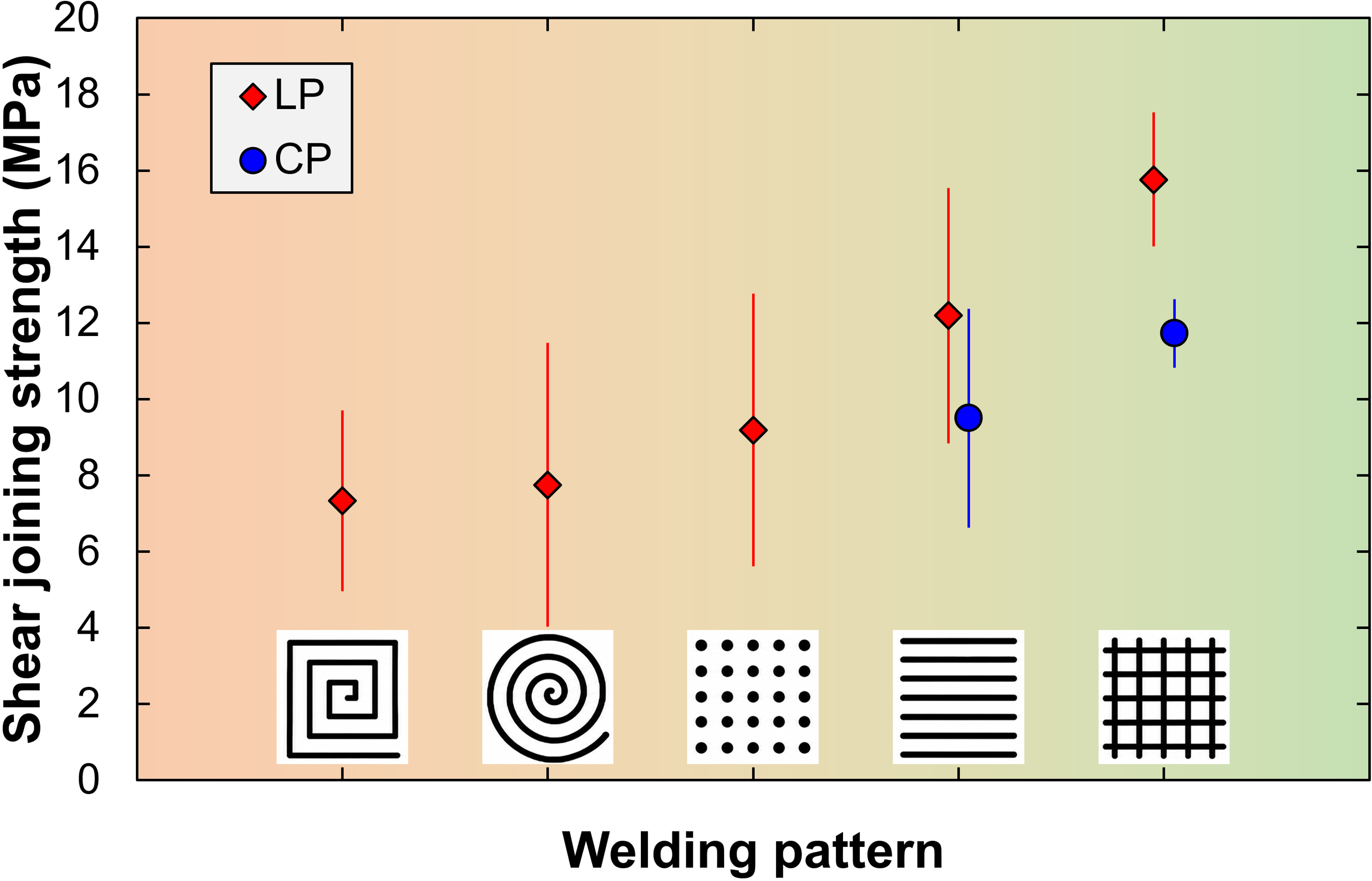}
\caption{\label{fig:Fig4} Evolution of the shear joining strength as a function of the welding pattern. The repeatability error bars indicate the standard deviation on the measured shear joining strength. The red and blue points correspond to linear polarization (LP) and circular polarization (CP), respectively. Pulse energy: $E_{\rm{in}}=6$~$\mu$J, focal position: $z = 380$~$\mu$m, metal roughness: $R_{q}=2.8$~nm.}
\end{figure}

To do so, we consider five different patterns: rectangular spiral, Archimedean spiral, point array, parallel lines, and grid.
For all these patterns, the welded area is similar (on the order of 9~mm$^{2}$, i.e., $\approx 36 \%$ of the 25~mm$^{2}$ available area corresponding to the silicon sample dimensions).
The results in terms of shear joining strength are summarized in Fig.~\ref{fig:Fig4}.
Both rectangular and Archimedean spiral patterns lead to a similar shear joining strength in the 7.0--8.0~MPa range.
The strength value is significantly increased for the point array pattern (9.2~MPa) and the line pattern (12.9~MPa).
Finally, the shear joining strength reaches 15.8~MPa for the grid pattern, consistent with recent observations in the glass-ceramic configuration \cite{Yu2026}.\\

An important parameter governing energy deposition in laser--matter interaction is the polarization, i.e., the direction of the electric field.
For the two patterns exhibiting the best performance (parallel lines and grid), we have tested linear and circular polarization, as shown with the red and blue data points in Fig.~\ref{fig:Fig4}.
For both patterns, circular polarization underperforms linear polarization, with $\approx 25\%$ decrease in the shear joining strength.
This is consistent with the improved energy deposition with linear polarization compared to circular polarization \cite{Shin2022}.
Notably, for the parallel line pattern, linear polarizations along and perpendicular to the writing direction were tested (not shown here).
The results are very similar for both configurations, with a shear joining strength difference $<8\%$, i.e., smaller than the error bars.

\subsubsection{Metal roughness}
\label{sssec:roughness}

\begin{figure}[!ht]
\centering
\includegraphics[width=\linewidth]{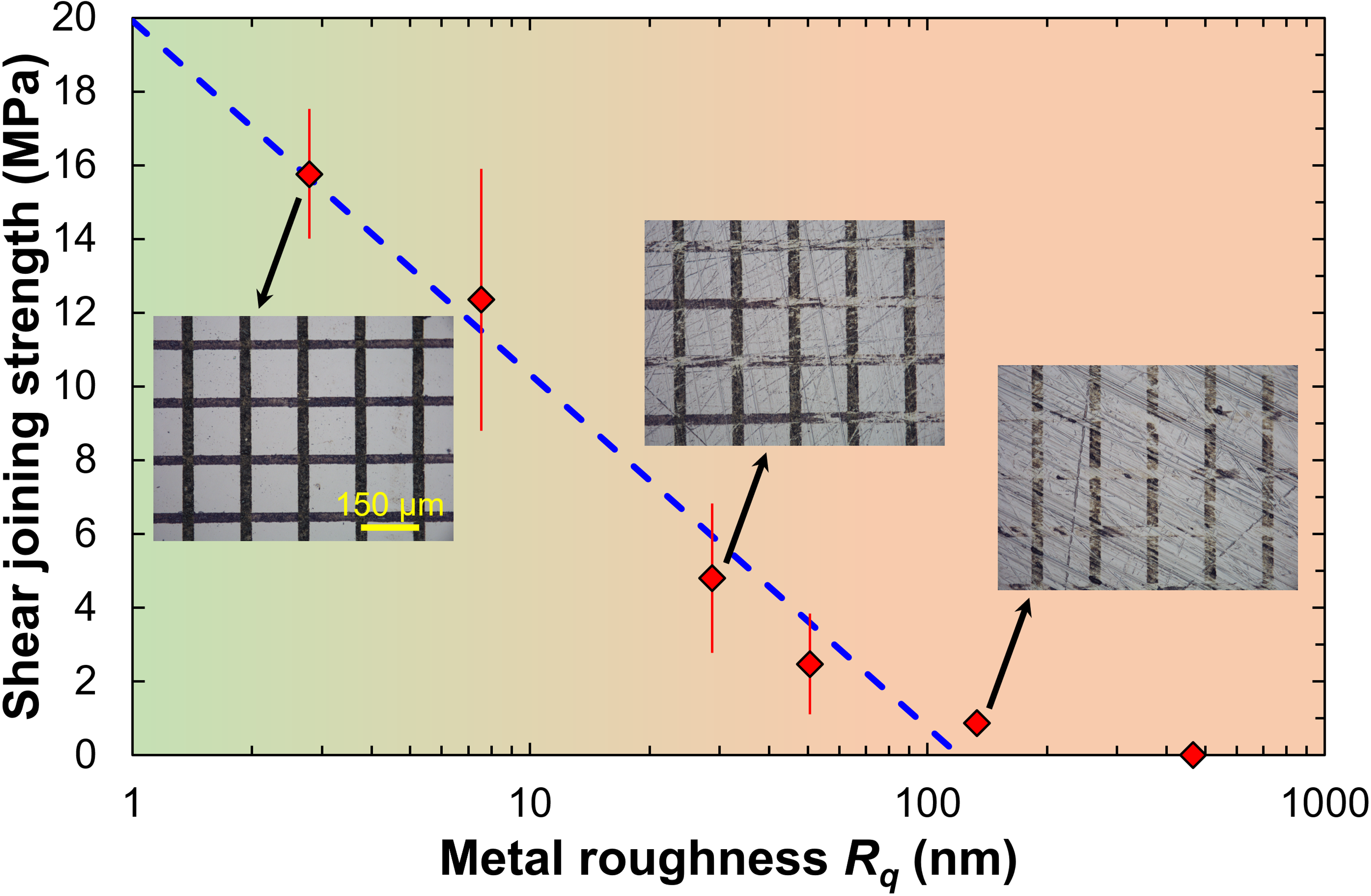}
\caption{\label{fig:Fig5} Evolution of the shear joining strength as a function of the root mean square (RMS) surface roughness $R_{q}$ of the metal sample. The repeatability error bars indicate the standard deviation on the measured shear joining strength. The blue dashed line is a logarithmic fit. The insets are optical micrographs of the welded area on the metal sample ($R_{q}=2.8$~nm, 28.7~nm, and 133.4~nm) after sample separation. The spatial scale applies to all images. Pulse energy: $E_{\rm{in}}=6$~$\mu$J, focal position: $z = 380$~$\mu$m, welding pattern: grid, laser polarization: linear.}
\end{figure}

In most experiments reported in this article, the silicon and metal samples are in optical contact during the welding.
The quality of the optical contact depends essentially on the sample roughness.
In this section, microelectronics-grade polished silicon samples have been joined to metal samples of various surface roughness values in the $R_{q}=2.8$--466~nm range.
The performance of the corresponding welds in terms of shear joining strength is displayed in Fig.~\ref{fig:Fig5}.
The results unambiguously show that silicon--metal laser welding performs better for smoother metal surfaces (best performance for $R_{q}=2.8$~nm), in good agreement with Refs.~\cite{Richter2015,Li2025}.
Interestingly, the shear joining strength scales logarithmically with the roughness (see the dashed blue line in Fig.~\ref{fig:Fig5}).
This implies that a substantial effort to improve the surface roughness will result in only a moderate improvement in welding performance.
In addition, as highlighted by the optical micrographs of the Kovar sample after sample separation (insets in Fig.~\ref{fig:Fig5}), the welding seam is more homogeneous for low $R_{q}$ values.
The presence of randomly distributed preexisting scratches strongly affects the generation of laser-induced modifications.\\

The trends in Fig.~\ref{fig:Fig5} can be ascribed to loss of optical contact for excessive $R_{q}$ values.
Indeed, the presence of an interfacial air gap has two major consequences.
First, Fresnel reflection at the silicon--air interface causes a loss of $\approx 30\%$ of the pulse energy before it reaches the metal.
Second, air is a near-perfect thermal insulator which may limit heat diffusion, and thus material transfer between silicon and metal, in turn hindering the formation of strong bonds between the two materials.
Beyond these aspects, Li \textit{et al.} recently proposed a qualitative scenario for laser joining across different interfacial gaps, supported by scanning electron microscopy, energy-dispersive X-ray spectroscopy, and strength measurements \cite{Li2025}.
This scenario describes a transition from smooth bonding for small gaps, to spallation-assisted metal eruption and mechanical interlocking for intermediate gaps, and finally to crack-dense, defective joints for larger gaps.
Although this scenario was proposed for sapphire--Invar laser welding, similar processes may also exist for silicon--Kovar laser welding.

\subsubsection{Comparison with traditional bonding techniques}
\label{sssec:othertechniques}

In industry, traditional techniques conventionally used to bond silicon to metal include adhesive bonding, eutectic bonding, and soldering.
In the present study, the optimized laser-welded joints exhibit a shear joining strength of 15.8~MPa, representing a substantial increase over the previously reported maximum of 2.2~MPa for silicon--metal laser welding \cite{Chambonneau2021a}.
Table~\ref{tab:othertechniques} compares this value with representative shear strengths reported for adhesive bonding, eutectic bonding, and soldering involving silicon.
The values were obtained for different material configurations and test geometries and are therefore intended as approximate technological benchmarks rather than strictly equivalent measurements.
The shear joining strength achieved by laser welding falls within the representative ranges reported for all three conventional bonding techniques.\\

\begin{table}[!ht]
\centering
\caption{Representative shear joining strengths reported for selected bonding techniques involving silicon.}
\label{tab:othertechniques}
\begin{adjustbox}{max width=\columnwidth}
\begin{tabular}{@{}llc@{}}
\toprule
\textbf{Bonding technique} & \textbf{Reported shear strength (MPa)} & \textbf{Reference}\\
\midrule
Laser welding & 15.8 & Present study\\
Adhesive bonding & 2.0--19.2 & \cite{Reuter2005,Stankovic2011,Keyvaninia2013,Bleiker2017}\\
Eutectic bonding & 4.2--20.4 & \cite{Jing2010,Abouie2012}\\
Soldering & 7.9--43.0 & \cite{Kolenak2016,Kolenak2021,Ding2022,Li2023,Xue2024}\\
\bottomrule
\end{tabular}
\end{adjustbox}
\end{table}

A common feature of adhesive bonding, soldering, and eutectic bonding is that the joint is mediated by an intentionally introduced bonding material, such as a polymer adhesive, a solder alloy, or a eutectic-forming metallic layer.
Although these approaches are well established, the intermediate material may introduce limitations under harsh-environment operation.
Polymer adhesives may undergo thermal degradation, aging, moisture uptake, or outgassing, whereas solders may soften, creep, or remelt when the temperature approaches their melting range.
Eutectic and soldered interfaces may also undergo microstructural evolution through diffusion and interfacial reactions during prolonged thermal exposure.
In addition, an intermediate layer introduces a third coefficient of thermal expansion into the assembly, which may generate additional thermomechanical stresses during temperature changes and compromise the integrity of the joint.
In contrast, a key advantage of direct laser welding is that the silicon--metal joint is formed without introducing an intermediate bonding layer.
This simplifies the material stack and avoids failure mechanisms directly associated with such added layers.
The thermal resistance and hermeticity of the silicon--metal joints produced by laser welding are examined in the following sections.\\

\subsection{Thermal resistance}
\label{ssec:temperature}

In Section~\ref{ssec:optimization}, the focal position, the welding pattern, the laser polarization as well as the metal roughness have been optimized.
For the optimized parameters resulting in a shear joining strength of 15.8~MPa (pulse energy: $E_{\rm{in}}=6$~$\mu$J, focal position: $z = 380$~$\mu$m, welding pattern: grid, laser polarization: linear, metal roughness: $R_{q}=2.8$~nm), the resilience to high-temperature environments is evaluated.
This step is critical for assessing the suitability of the proposed technique for operation in harsh environments.
Assessing the high-temperature stability of the fabricated silicon--metal bond is relevant for numerous harsh-environment applications, including jet- and rocket-engine diagnostics, nuclear instrumentation, and metallurgical sensors.\\

To assess weld lifetime under thermal loading, analogous studies on the thermal resistance of laser-joined materials have relied on thermal cycling with relatively low maximum temperatures ($<200$\,\degree C) \cite{Carter2017,Morawska2024,Dzipalski2024,Huo2025}.
The weld survival results strongly depend on the difference in the linear coefficient of thermal expansion $\alpha$ between the two joined materials.
This conclusion is reasonable as the laser-created bonds between two materials expanding differently with a temperature change are likely to be weakened.
For instance, in the case of fused silica joined to aluminum, the ratio between the coefficients of thermal expansion is $\alpha_{\rm{Al}}/\alpha_{\rm{SiO_{2}}}=47$ \cite{Carter2017}.
In contrast, as highlighted in Table~\ref{tab:thermalproperties} where the thermal properties of silicon and Kovar are summarized, the ratio between the coefficients of thermal expansion at room temperature $\alpha_{\rm{Kovar}}/\alpha_{\rm{Si}}=1.9$ is drastically lower.
Moreover, when the temperature is increased in the 100--700\,\degree C range, the ratio $\alpha_{\rm{Kovar}}/\alpha_{\rm{Si}}$ lies in the 1.5--2.4 range \cite{Watanabe2004,LamineriesMATTHEY}, i.e., of the same order of magnitude as at room temperature. This suggests that the silicon--metal welds produced in our study should better withstand high-temperature exposure than fused silica--aluminum welds.\\

\begin{table}[!ht]
\centering
\caption{Thermal properties of silicon and Kovar at room temperature.}
\label{tab:thermalproperties}
\begin{adjustbox}{max width=\columnwidth}
\begin{tabular}{@{}lcc@{}}
\toprule
\textbf{Material} & \textbf{silicon} & \textbf{Kovar}\\
\midrule
Melting point (\degree C) & 1414 \cite{Yamaguchi2002} & 1450 \cite{Chen2023}\\
Coefficient of thermal expansion ($10^{-6}/$\degree C) & 2.6 \cite{Watanabe2004} & 5.0 \cite{Zanchetta1995}\\
Thermal diffusivity (mm$^{2}$/s) & 86 \cite{Shanks1963} & 4.2 \cite{LamineriesMATTHEY}\\
\bottomrule
\end{tabular}
\end{adjustbox}
\end{table}

\begin{figure}[!ht]
\centering
\includegraphics[width=\linewidth]{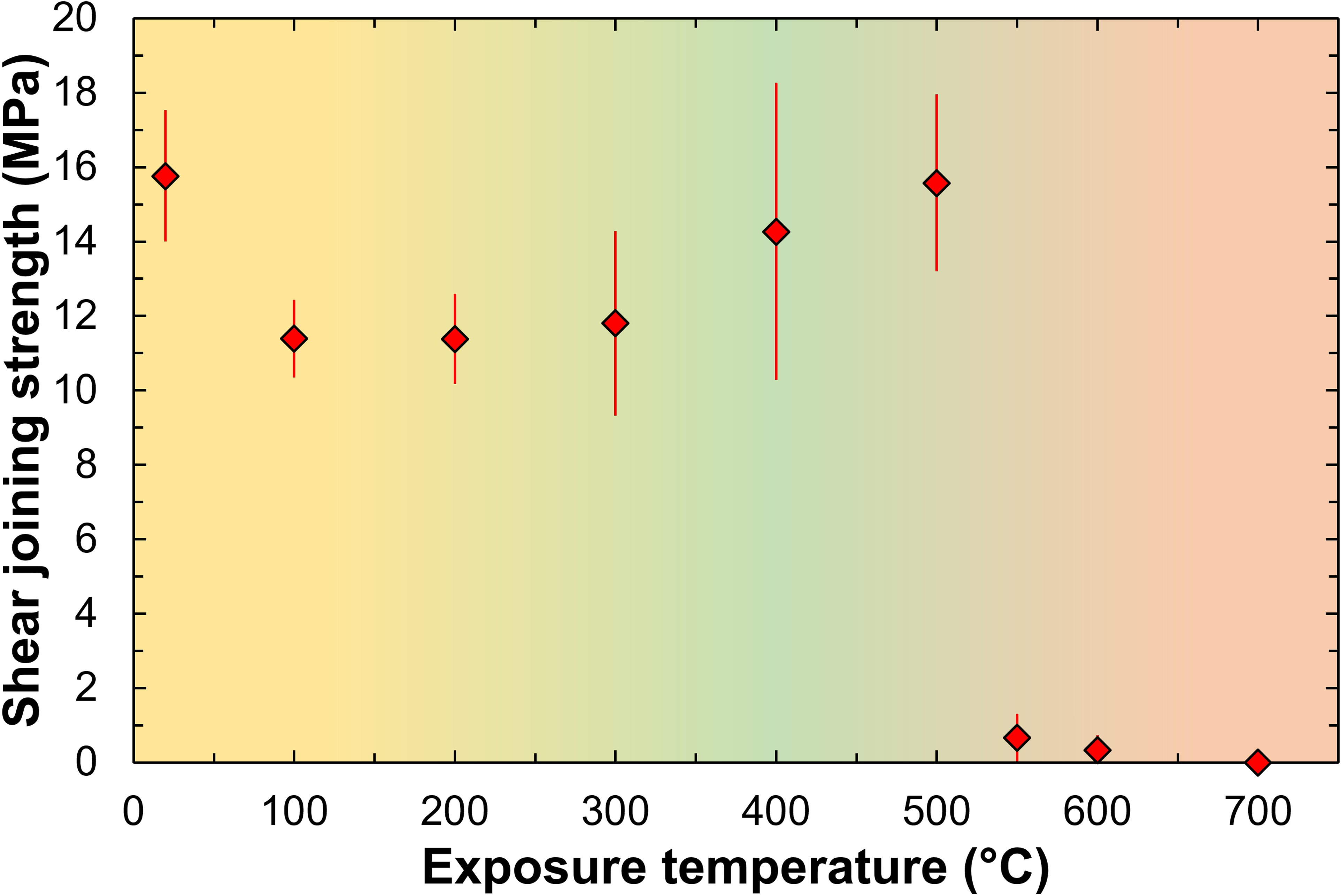}
\caption{\label{fig:Fig6} Evolution of the shear joining strength as a function of the exposure temperature. The repeatability error bars indicate the standard deviation on the measured shear joining strength. Pulse energy: $E_{\rm{in}}=6$~$\mu$J, focal position: $z = 380$~$\mu$m, welding pattern: grid, laser polarization: linear, metal roughness: $R_{q}=2.8$~nm, duration in furnace: one hour.}
\end{figure}

To establish the thermal resistance, instead of thermal cycling, we rely on a one-hour exposure to a set temperature in a furnace followed by a one-hour cooling period to room temperature.
The corresponding results are shown in Fig.~\ref{fig:Fig6} in terms of shear joining strength for various exposure temperatures.
Taking measurements of the bonding strength at room temperature ($T=20$\,\degree C) as a reference, the temperature dependence can be divided into three regimes.
First, increasing the temperature up to $T=300$\,\degree C leads to a decrease in shear joining strength by $\approx 27\%$.
This may be attributed to the thermal expansion mismatch between silicon and Kovar, which can activate precursor defects, microcracks, or weak interfacial regions formed during laser welding, as suggested by optical microscopy observations of the Kovar surface after thermal exposure at different temperatures (see Appendix~B).
In the second regime, for higher temperatures ($T=300$--500\,\degree C), the shear joining strength progressively returns to values comparable to the reference case at room temperature.
In this intermediate temperature range, stress relief and annealing effects may partially compensate for the aforementioned thermally induced damage.
The interfacial region may become mechanically more stable after heat treatment, so the final room-temperature strength recovers.
Finally, in the third regime, for $T\ge 550$\,\degree C, the shear joining strength abruptly drops to sub-MPa values, which indicates that the silicon and metal parts are no longer effectively joined after high-temperature exposure.
At these temperatures, accumulated thermomechanical stress, interfacial cracking, metal oxidation, and oxide growth may contribute to bond failure.
To conclude, the silicon--metal laser joints survive a one-hour exposure up to $T = 500$\,\degree C with residual room-temperature shear joining strength comparable to the as-welded reference, but they lose practical joining strength after exposure to $T > 500$\,\degree C.
Thus, these silicon--metal welds are compatible with high-temperature applications.\\

\subsection{Hermeticity}
\label{ssec:hermeticity}

The second aspect considered concerns high-vacuum operation and hermetic-sealing requirements.
Assessing the leak rate of silicon--metal welds is particularly relevant for applications in which long-term isolation from gases, moisture, or aggressive environments is critical, including space and aerospace electronics, implantable medical devices, particle-physics detectors, vacuum-packaged micro-electromechanical systems (MEMS), quantum photonic devices, and underwater packages.\\

\begin{figure}[!ht]
\centering
\includegraphics[width=\linewidth]{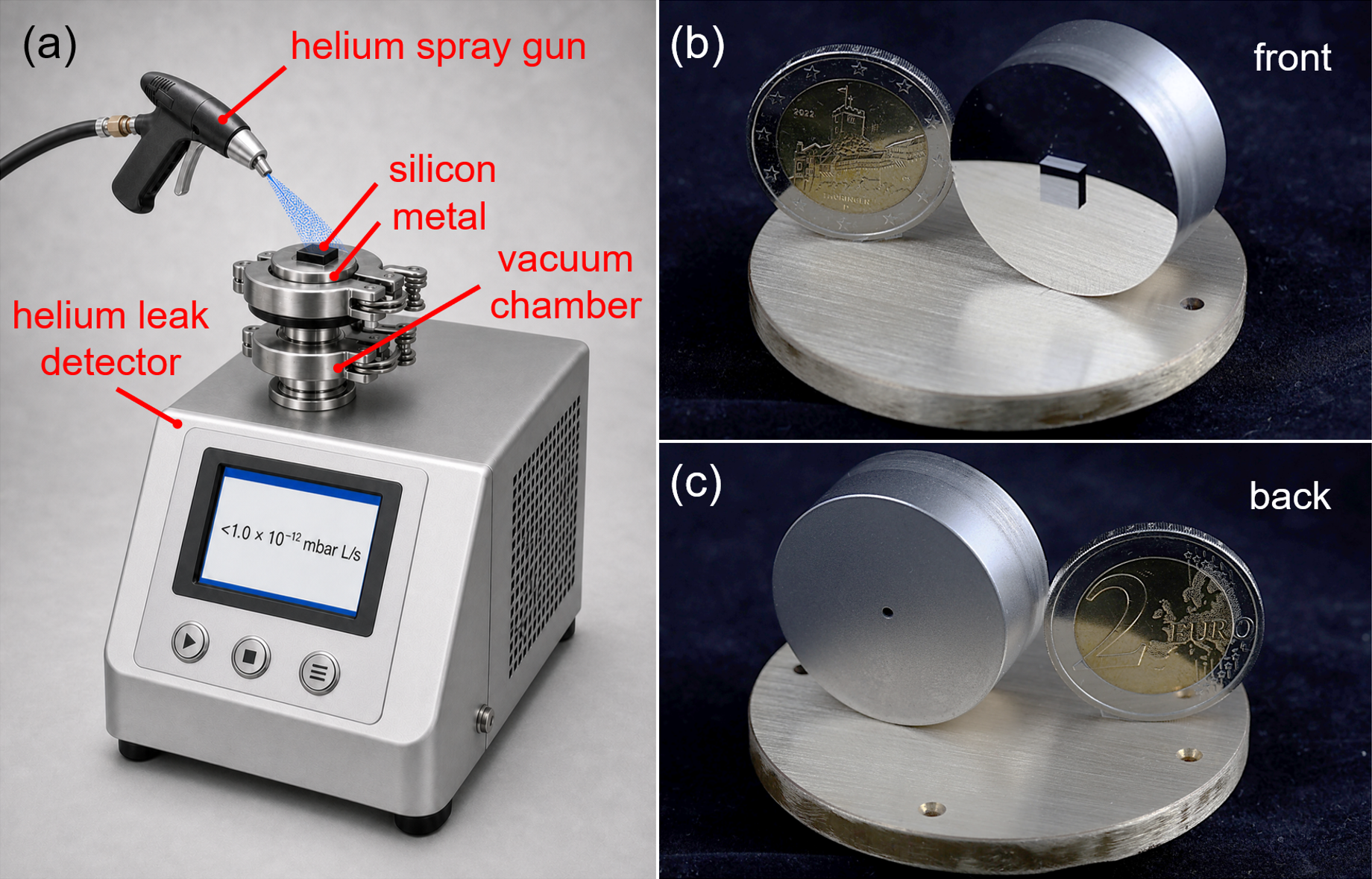}
\caption{\label{fig:Fig7} Hermetic silicon--metal welds. (a) Schematic of the experimental setup for hermeticity measurements. (b) Front and (c) back side of the tested silicon--metal welds. Pulse energy: $E_{\rm{in}}=6$~$\mu$J, focal position: $z = 380$~$\mu$m, welding pattern: grid, laser polarization: linear, metal roughness: $R_{q}=2.8$~nm.}
\end{figure}

As illustrated in Fig.~\ref{fig:Fig7}(a), the hermeticity of the silicon--metal joints was evaluated using helium leak testing, with each joint sealing a vacuum chamber connected to a mass spectrometer while helium was sprayed onto its opposite side.
Helium was selected as a highly sensitive tracer gas because its low molar mass makes it particularly effective for detecting small geometrical leak paths, especially in the molecular-flow regime.
For these tests, a 1-mm diameter cylindrical hole is drilled in the metal plate, with the silicon sample fully covering one end of the hole [see Fig.~\ref{fig:Fig7}(b) and (c)].
This implies that imperfect contact would allow helium atoms to pass through the hole and eventually be detected by the mass spectrometer.\\

However, this is not the case here, and the silicon--metal joints are in fact leak-proof.
All five tested joints showed no helium leak above the detection limit of $Q=1.0 \times 10^{-12}$~mbar~L/s, defined as $Q=P dV/dt$, where $P$ is the pressure and $dV/dt$ is the equivalent volumetric flow rate of the leaked helium.
The measured upper bound thus corresponds to only $\approx 30$~nanoliters of helium per year at atmospheric pressure, making it completely negligible for most vacuum applications.
The leak rate $<1.0 \times 10^{-12}$~mbar~L/s therefore indicates ultra-high hermeticity within the detection limit of the measurements.\\

To benchmark this performance, leak-rate values reported in the literature for selected laser-joined material configurations are compared with the present silicon--metal welds in Table~\ref{tab:leakrate}.
Because these studies involve different material systems, geometries, test gases, and measurement protocols, the comparison should be interpreted as an order-of-magnitude benchmark rather than as a strict ranking.
Within this limitation, the upper bound $Q=1.0 \times 10^{-12}$~mbar~L/s reported here compares favorably with previous laser-joined ceramic--ceramic, glass--glass, glass--silicon, and glass-ceramic--glass-ceramic systems \cite{Penilla2019,Su2025,Wang2025b,Yu2026}.
It is at least one order of magnitude lower than the best literature value listed in Table~\ref{tab:leakrate}, and up to six orders of magnitude lower than the value reported for glass-ceramic laser sealing.

\begin{table}[!ht]
\centering
\caption{Leak rates reported for selected laser-joined material systems.}
\label{tab:leakrate}
\begin{adjustbox}{max width=\columnwidth}
\begin{tabular}{@{}llc@{}}
\toprule
\textbf{Reference} & \textbf{Laser-joined materials} & \textbf{Leak rate (mbar~L/s)}\\
\midrule
Present study & silicon--Kovar & $<1.0 \times 10^{-12}$\\
Penilla \textit{et al.} \cite{Penilla2019} & ceramic--ceramic & $1.3 \times 10^{-10}$\\
Su \textit{et al.} \cite{Su2025} & glass--glass & $1.0 \times 10^{-11}$\\
Wang \textit{et al.} \cite{Wang2025b} & glass--silicon & $7.4 \times 10^{-9}$\\
Yu \textit{et al.} \cite{Yu2026} & glass-ceramic--glass-ceramic & $1.2 \times 10^{-6}$\\
\bottomrule
\end{tabular}
\end{adjustbox}
\end{table}

\section*{Conclusion}
\label{sec:conclusion}

To summarize, successful silicon--metal laser welding exhibiting shear joining strengths up to $15.8$~MPa has been demonstrated.
To reach such high strengths, we optimized multiple laser and material parameters.
At the selected input pulse energy of $E_{\rm{in}}=6$~$\mu$J, for which a $100\%$ modification probability is reached, the focus is positioned at $z = 380$~$\mu$m below the silicon--metal interface, ensuring optimal energy deposition at the interface.
The highest strength values are obtained for a grid pattern under linear laser polarization.
We have shown that minimizing the metal roughness $R_{q}$ down to 2.8~nm ensures an excellent contact between the two workpieces throughout welding, and thus the best welding performance.
After this optimization step, high-temperature exposure tests revealed that the silicon--metal welds are still operational for temperatures up to 500\,\degree C.
Finally, hermeticity measurements showed a leak rate $<1.0 \times 10^{-12}$~mbar~L/s, highlighting that the silicon--metal welds are hermetically sealed.\\

These results in terms of shear joining strength offer an alternative to traditional joining methods (e.g., adhesive bonding, eutectic bonding, soldering).
As silicon and metals are the cornerstones of microelectronics, the proposed welding technique should find multiple applications in this field.
Beyond standard microelectronics, high-temperature-resistant and hermetic silicon--metal joints may be of utmost importance in harsh-environment applications such as aerospace.

\section*{Data availability}

The data that support the results of this article are available within the manuscript.

\section*{Funding}

This research has received financial support from:
\begin{itemize}
    \item the German Federal Ministry of Research, Technology and Space [Bundesministerium für Forschung, Technologie und Raumfahrt (BMFTR)] through the RUBIN-UKPi\~no project (Grants no.~03RU2U033H, and 03RU2U032F).
    \item the German Research Foundation [Deutsche Forschungsgemeinschaft (DFG)], through the Silabus (Grant no.~530105422), and the Inseption (Grant no.~545531713) projects.
\end{itemize}

\section*{Acknowledgments}

The authors gratefully acknowledge the technical support of Markus~Walther (Institute of Applied Physics, Jena, Germany) for cutting samples, Eric~Schadow (Fraunhofer IOF, Jena, Germany) for roughness measurements, and Qingfeng~Li (Thorlabs GmbH, Bergkirchen, Germany) for designing the clamping system.

\section*{Competing interests}

The authors declare no competing interests.

\section*{Author contributions}

J.Y. performed the welding experiments conceived by M.C. and carried out the strength measurements. M.B. and M.C. designed and implemented the experimental setup. H.P.K. provided technical assistance with polishing and with the clamping system. C.S. performed the hermeticity measurements. S.N. and M.C. supervised the research and acquired funding. M.C. prepared the manuscript with contributions from all authors.

\setcounter{equation}{0}
\setcounter{figure}{0}
\setcounter{table}{0}
\setcounter{section}{0}
\renewcommand{\theequation}{S\arabic{equation}}
\renewcommand{\thesection}{S\arabic{section}}
\renewcommand{\thesubsection}{S\arabic{section}.\arabic{subsection}}
\renewcommand{\thefigure}{S\arabic{figure}}
\renewcommand{\thetable}{S\arabic{table}}

\section*{Appendix~A. Modification probability}

A prerequisite for a continuous and homogeneous welding pattern is that irradiation under the selected processing conditions reliably produces interfacial material modification.
In other words, the modification probability has to be close to $100\%$.
This probability is defined as the ratio between the number of modified sites and the total number of tested sites for a given pulse energy.
Modification probabilities significantly lower than $100\%$ may lead to decreased mechanical performance of the welds---or no welding at all.\\

\begin{figure}[!ht]
\centering
\includegraphics[width=\linewidth]{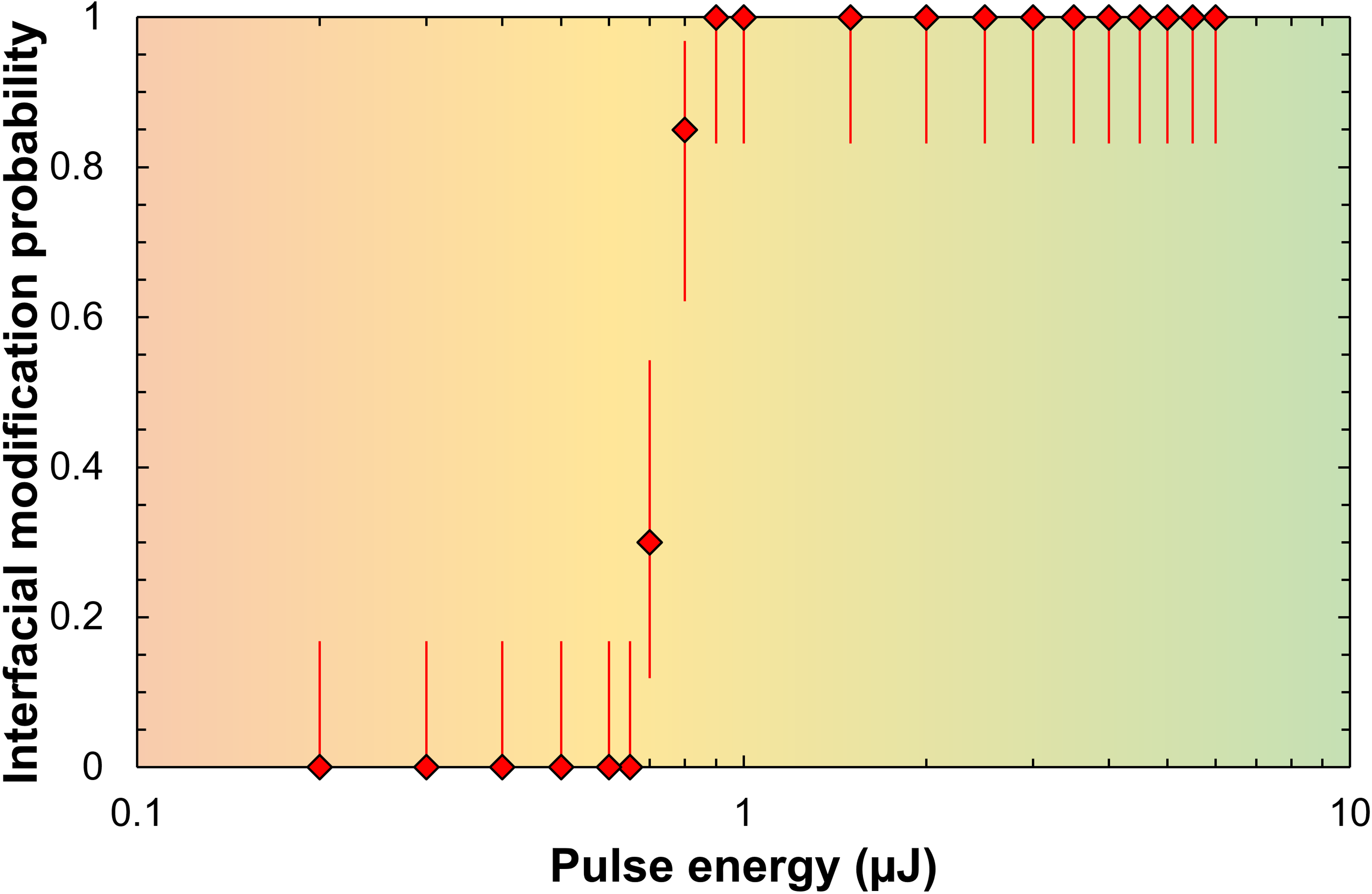}
\caption{\label{fig:FigS1} Evolution of the interfacial modification probability as a function of the input pulse energy $E_{\rm{in}}$. The focus is positioned at $z=380$~$\mu$m. The error bars correspond to $95\%$ Clopper--Pearson binomial confidence bounds, calculated from the number of modifications among 20 fresh sites tested at each pulse energy and spatially separated by 150~$\mu$m. To ease modification detection, 6250 pulses were applied per site (S-on-1 procedure governed by ISO 21254-2 \cite{standarddamage}).}
\end{figure}

For constant pulse duration and focal spot size, modification occurrence mainly depends on the pulse energy.
To investigate how the interfacial modification probability evolves with the pulse energy, we exploit our customized infrared microscope for \textit{in-situ} modification detection at the silicon--metal interface; the resulting observations are consistent with \textit{ex-situ} optical microscopy observations [Fig.~\ref{fig:Fig2}(b) and (c)].
The results obtained at $z=380$~$\mu$m, the focal position yielding the best welding performance (see Fig.~\ref{fig:Fig3}), are displayed in Fig.~\ref{fig:FigS1}.
The interfacial modification probability typically follows a sigmoid-like dependence on the input pulse energy, from $0\%$ to $100\%$.
This is consistent with the relatively long pulse duration (620~ps), in contrast with single-digit picosecond pulse durations where extreme propagation nonlinearities lead to a non-monotonic behavior \cite{Chambonneau2019,Das2020,Blothe2025}.
The modification threshold is estimated to be $E_{\rm{th}}\approx 675 \pm 25$~nJ.
The interfacial modification probability increases with the pulse energy, until it reaches $100\%$ for an input pulse energy of $E_{\rm{in}}\approx 850 \pm 50$~nJ.
This implies that for $E_{\rm{in}} > 900$~nJ, the welding pattern should be homogeneous.
In all experiments shown in this study, we have chosen $E_{\rm{in}}=6$~$\mu$J, as this is well above this threshold and maximizes the energy reaching the interface \cite{Chambonneau2021a}.\\

\section*{Appendix~B. Temperature-dependent Kovar morphology}

\begin{figure}[!ht]
\centering
\includegraphics[width=\linewidth]{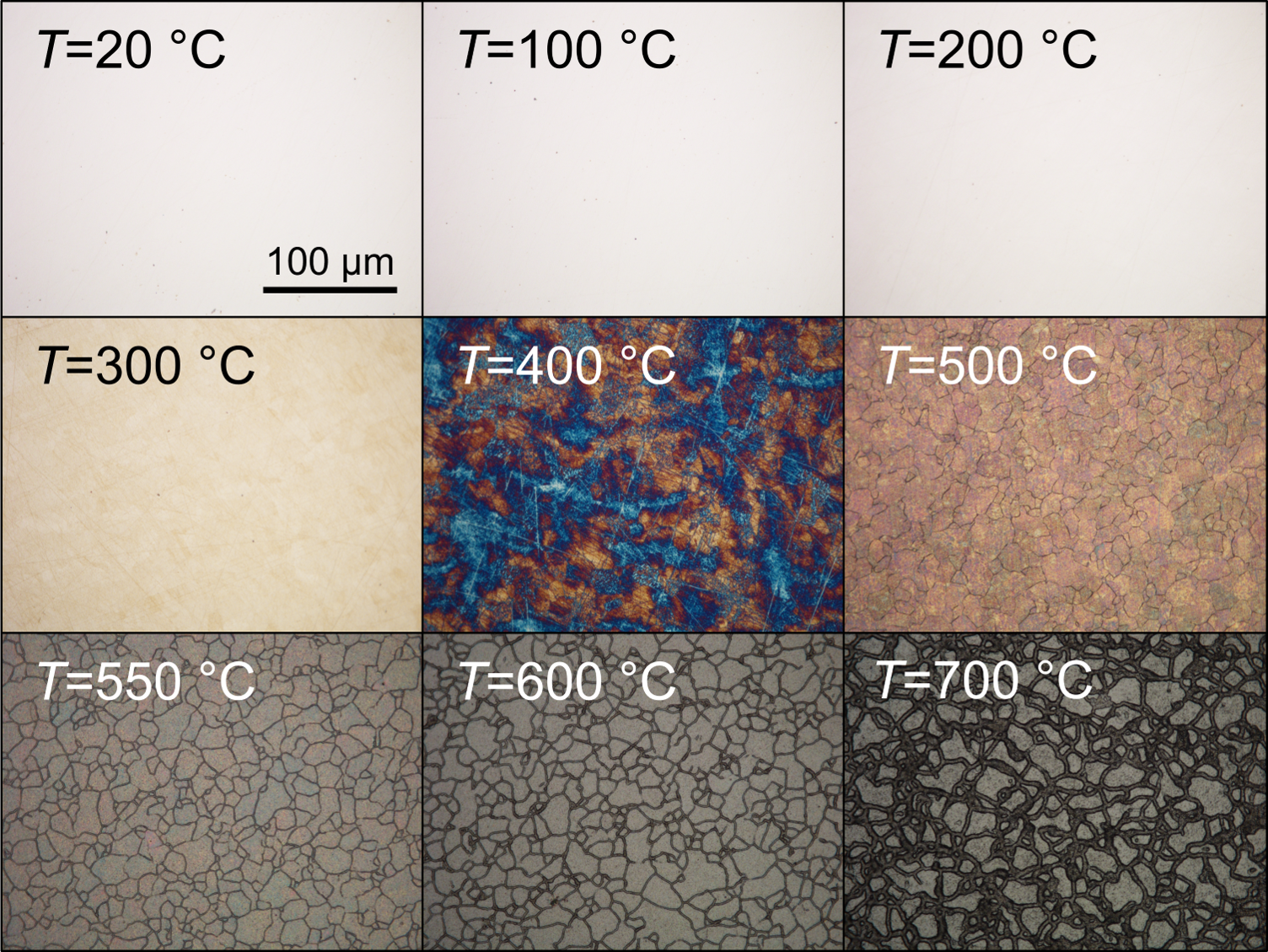}
\caption{\label{fig:FigS2} Optical micrographs of unirradiated Kovar samples after thermal exposure at the indicated temperatures $T$. The scale applies to all images.}
\end{figure}

After thermal exposure (see Section~\ref{ssec:temperature}), the Kovar samples were inspected under optical microscopy.
The corresponding unirradiated areas are displayed in Fig.~\ref{fig:FigS2}.
For temperatures $T \le 200$\,\degree C, the surface appearance is the same as at room temperature.
However, a noticeable color change is observed on the surface for samples heated to $300 \le T < 500$\,\degree C, which can be explained by metal oxidation.
The oxide layer thickness increases with temperature.
The striking feature in Fig.~\ref{fig:FigS2} is the appearance of a crack network on the oxide layer for $T \ge 500$\,\degree C.
The cracks are barely noticeable at $T = 500$\,\degree C, which is consistent with the high performance of the silicon--Kovar joints after exposure to this temperature (see Fig.~\ref{fig:Fig6}).
However, when the temperature is increased ($550 \le T \le 700$\,\degree C), the crack width increases substantially.
At these temperatures, the shear joining strength of the silicon--Kovar joints decreases dramatically to $<1$~MPa.
We therefore conclude that crack formation is among the main factors responsible for the decreased mechanical performance at temperatures $T \ge 550$\,\degree C.\\

\bibliographystyle{unsrtnat}
{\footnotesize
\bibliography{refs}

@article{Sahoo2020,
   author = {Pankaj K. Sahoo and Tao Feng and Jie Qiao},
   doi = {10.1364/oe.402493},
   issn = {10944087},
   issue = {21},
   journal = {Optics Express},
   month = {10},
   pages = {31103},
   pmid = {33115092},
   publisher = {Optica Publishing Group},
   title = {Dynamic pulse propagation modelling for predictive femtosecond-laser-microbonding of transparent materials},
   volume = {28},
   url = {https://doi.org/10.1364/OE.402493},
   year = {2020}
}

@article{Chambonneau2023,
   author = {Maxime Chambonneau and Qingfeng Li and Markus Blothe and Stree Vithya Arumugam and Stefan Nolte},
   doi = {10.1002/adpr.202200300},
   issn = {2699-9293},
   issue = {5},
   journal = {Advanced Photonics Research},
   month = {5},
   publisher = {Wiley},
   title = {Ultrafast Laser Welding of Silicon},
   volume = {4},
   url = {https://doi.org/10.1002/adpr.202200300},
   year = {2023}
}

@article{Chambonneau2026,
   author = {Maxime Chambonneau and Markus Blothe and Vladimir Yu. Fedorov and Isaure de Kernier and Stelios Tzortzakis and Stefan Nolte},
   doi = {10.1038/s41467-026-69530-w},
   issn = {2041-1723},
   issue = {1},
   journal = {Nature Communications},
   month = {2},
   pages = {1701},
   publisher = {Nature Research},
   title = {Extreme optical nonlinearities unveiled by ultrafast laser filamentation in semiconductors},
   volume = {17},
   url = {https://www.nature.com/articles/s41467-026-69530-w},
   year = {2026}
}

@article{Chambonneau2021b,
   author = {Maxime Chambonneau and David Grojo and Onur Tokel and Fatih Ömer Ilday and Stelios Tzortzakis and Stefan Nolte},
   doi = {10.1002/lpor.202100140},
   issn = {18638899},
   issue = {11},
   journal = {Laser $\&$ Photonics Reviews},
   month = {11},
   publisher = {John Wiley and Sons Inc},
   title = {In-Volume Laser Direct Writing of Silicon—Challenges and Opportunities},
   volume = {15},
   url = {https://doi.org/10.1002/lpor.202100140},
   year = {2021}
}

@article{Chambonneau2021a,
   author = {Maxime Chambonneau and Qingfeng Li and Vladimir Yu. Fedorov and Markus Blothe and Kay Schaarschmidt and Martin Lorenz and Stelios Tzortzakis and Stefan Nolte},
   doi = {10.1002/lpor.202000433},
   issn = {1863-8880},
   issue = {2},
   journal = {Laser $\&$ Photonics Reviews},
   month = {2},
   publisher = {Wiley-VCH Verlag},
   title = {Taming Ultrafast Laser Filaments for Optimized Semiconductor–Metal Welding},
   volume = {15},
   url = {https://onlinelibrary.wiley.com/doi/10.1002/lpor.202000433},
   year = {2021}
}

@article{Li2025b,
   author = {Qingfeng Li and Fei Luo and Gabor Matthäus and David Sohr and Stefan Nolte},
   doi = {10.3390/nano15161215},
   issn = {2079-4991},
   issue = {16},
   journal = {Nanomaterials},
   month = {8},
   pages = {1215},
   publisher = {Multidisciplinary Digital Publishing Institute (MDPI)},
   title = {Direct Glass-to-Metal Welding by Femtosecond Laser Pulse Bursts: II, Enhancing the Weld Between Glass and Polished Metal Surfaces},
   volume = {15},
   url = {https://www.mdpi.com/2079-4991/15/16/1215},
   year = {2025}
}

@article{Li2025a,
   author = {Qingfeng Li and Gabor Matthäus and David Sohr and Stefan Nolte},
   doi = {10.3390/nano15151202},
   issn = {2079-4991},
   issue = {15},
   journal = {Nanomaterials},
   month = {8},
   pages = {1202},
   publisher = {Multidisciplinary Digital Publishing Institute (MDPI)},
   title = {Direct Glass-to-Metal Welding by Femtosecond Laser Pulse Bursts: I, Conditions for Successful Welding with a Gap},
   volume = {15},
   url = {https://www.mdpi.com/2079-4991/15/15/1202},
   year = {2025}
}

@article{Bristow2007,
   author = {Alan D. Bristow and Nir Rotenberg and Henry M. van Driel},
   doi = {10.1063/1.2737359},
   issn = {0003-6951},
   issue = {19},
   journal = {Applied Physics Letters},
   month = {5},
   title = {Two-photon absorption and Kerr coefficients of silicon for 850--2200 nm},
   volume = {90},
   url = {https://pubs.aip.org/apl/article/90/19/191104/327545/Two-photon-absorption-and-Kerr-coefficients-of},
   year = {2007}
}

@article{Sundaram2002,
   author = {S. K. Sundaram and E. Mazur},
   doi = {10.1038/nmat767},
   issn = {1476-1122},
   issue = {4},
   journal = {Nature Materials},
   month = {12},
   pages = {217-224},
   title = {Inducing and probing non-thermal transitions in semiconductors using femtosecond laser pulses},
   volume = {1},
   url = {https://www.nature.com/articles/nmat767},
   year = {2002}
}

@article{Gattass2008,
   author = {Rafael R. Gattass and Eric Mazur},
   doi = {10.1038/nphoton.2008.47},
   issn = {1749-4885},
   issue = {4},
   journal = {Nature Photonics},
   month = {4},
   pages = {219-225},
   title = {Femtosecond laser micromachining in transparent materials},
   volume = {2},
   url = {https://www.nature.com/articles/nphoton.2008.47},
   year = {2008}
}

@article{Li2025,
   author = {Nan Li and Yu Wang and Yitong Chen and Qingwei Zhang and Hongpeng Xiao and Yuxuan Li and Zhe Lin and Shanglu Yang},
   doi = {10.1364/OE.578149},
   issn = {1094-4087},
   issue = {23},
   journal = {Optics Express},
   month = {11},
   pages = {49661},
   pmid = {41414351},
   publisher = {Optica Publishing Group},
   title = {Ultrafast laser direct welding of sapphire and Invar under non-optical contact conditions with white-light-interferometric gap measurement},
   volume = {33},
   url = {https://opg.optica.org/abstract.cfm?URI=oe-33-23-49661},
   year = {2025}
}

@article{Yu2026,
   author = {Manduo Yu and Jiachang Deng and Liang Wang and Jun Luo and Jiadong Liang and Feng Zhang and Yu Zhang and Wei Zhang and Jiang Wang and Guodong Zhang and Guanghua Cheng},
   doi = {10.1016/j.optlastec.2025.114384},
   issn = {00303992},
   journal = {Optics $\&$ Laser Technology},
   month = {2},
   pages = {114384},
   publisher = {Elsevier Ltd},
   title = {Direct microwelding and sealing of glass-ceramics with burst-mode ultrafast laser},
   volume = {194},
   url = {https://linkinghub.elsevier.com/retrieve/pii/S0030399225019759},
   year = {2026}
}

@article{Richter2015,
   author = {S. Richter and F. Zimmermann and R. Eberhardt and A. Tünnermann and S. Nolte},
   doi = {10.1007/s00339-015-9377-8},
   issn = {0947-8396},
   issue = {1},
   journal = {Applied Physics A},
   month = {10},
   pages = {1-9},
   publisher = {Springer Verlag},
   title = {Toward laser welding of glasses without optical contacting},
   volume = {121},
   url = {http://link.springer.com/10.1007/s00339-015-9377-8},
   year = {2015}
}

@article{Das2020,
   author = {Amlan Das and Andong Wang and Olivier Uteza and David Grojo},
   doi = {10.1364/OE.398984},
   issn = {1094-4087},
   issue = {18},
   journal = {Optics Express},
   month = {8},
   pages = {26623},
   pmid = {32906932},
   publisher = {Optica Publishing Group},
   title = {Pulse-duration dependence of laser-induced modifications inside silicon},
   volume = {28},
   url = {https://opg.optica.org/abstract.cfm?URI=oe-28-18-26623},
   year = {2020}
}

@article{Chambonneau2019,
   author = {M. Chambonneau and L. Lavoute and D. Gaponov and V.Y. Fedorov and A. Hideur and S. Février and S. Tzortzakis and O. Utéza and D. Grojo},
   doi = {10.1103/PhysRevApplied.12.024009},
   issn = {2331-7019},
   issue = {2},
   journal = {Physical Review Applied},
   month = {8},
   pages = {024009},
   publisher = {American Physical Society},
   title = {Competing Nonlinear Delocalization of Light for Laser Inscription Inside Silicon with a 2-$\mu$m Picosecond Laser},
   volume = {12},
   url = {https://link.aps.org/doi/10.1103/PhysRevApplied.12.024009},
   year = {2019}
}

@article{Blothe2025,
   author = {Markus Blothe and Jesvin Joseph and Maxime Chambonneau and Stefan Nolte},
   doi = {10.1364/OL.575250},
   issn = {0146-9592},
   issue = {21},
   journal = {Optics Letters},
   month = {11},
   pages = {6650},
   pmid = {41185212},
   publisher = {Optica Publishing Group},
   title = {Internal silicon laser processing with picosecond double pulses},
   volume = {50},
   url = {https://opg.optica.org/abstract.cfm?URI=ol-50-21-6650},
   year = {2025}
}

@inproceedings{Dzipalski2024,
   author = {Adrian Dzipalski and Owen McGann and Richard M. Carter and M.J. Daniel Esser and Duncan P. Hand},
   doi = {10.1117/12.3031639},
   editor = {Chantal Andraud and Roberto Zamboni and Luana Persano and Andrea Camposeo},
   isbn = {9781510681187},
   issn = {1996756X},
   booktitle = {Advanced Materials, Biomaterials, and Manufacturing Technologies for Security and Defence II},
   month = {11},
   pages = {13},
   publisher = {SPIE},
   title = {Ultrashort pulsed laser welding of co-doped Er,Yb:Phosphate laser glass and Nd:YAG laser crystals to structural materials for robust mechanical mounting and thermal management},
   url = {https://doi.org/10.1117/12.3031639},
   year = {2024}
}

@article{Carter2017,
   author = {Richard M. Carter and Michael Troughton and Jianyong Chen and Ian Elder and Robert R. Thomson and M. J. Daniel Esser and Robert A. Lamb and Duncan P. Hand},
   doi = {10.1364/AO.56.004873},
   issn = {0003-6935},
   issue = {16},
   journal = {Applied Optics},
   month = {6},
   pages = {4873},
   pmid = {29047628},
   publisher = {Optica Publishing Group},
   title = {Towards industrial ultrafast laser microwelding: SiO$_2$ and BK7 to aluminum alloy},
   volume = {56},
   url = {https://opg.optica.org/abstract.cfm?URI=ao-56-16-4873},
   year = {2017}
}

@article{Morawska2024,
   author = {Paulina O. Morawska and Adrian Dzipalski and Tara Van Abeelen and Peter E. MacKay and Richard M. Carter and M. J. Daniel Esser and Duncan P. Hand},
   doi = {10.1364/OME.535427},
   issn = {2159-3930},
   issue = {11},
   journal = {Optical Materials Express},
   month = {11},
   pages = {2588},
   publisher = {Optica Publishing Group},
   title = {Industrial picosecond pulse laser welding of stainless-steel to quartz for optical applications},
   volume = {14},
   url = {https://opg.optica.org/abstract.cfm?URI=ome-14-11-2588},
   year = {2024}
}

@article{Huo2025,
   author = {Jingyu Huo and Zirong Zeng and Jinhui Yuan and Minghuo Luo and Aiping Luo and Jiaming Li and Huan Yang and Nan Zhao and Qingmao Zhang},
   doi = {10.1016/j.optlastec.2024.111804},
   issn = {00303992},
   journal = {Optics $\&$ Laser Technology},
   month = {2},
   pages = {111804},
   publisher = {Elsevier Ltd},
   title = {Welding between rough copper foil and silica glass using green femtosecond laser},
   volume = {181},
   url = {https://linkinghub.elsevier.com/retrieve/pii/S0030399224012623},
   year = {2025}
}

@article{Tamaki2005,
   author = {Takayuki Tamaki and Wataru Watanabe and Junji Nishii and Kazuyoshi Itoh},
   doi = {10.1143/JJAP.44.L687},
   issn = {0021-4922},
   issue = {5L},
   journal = {Japanese Journal of Applied Physics},
   month = {5},
   pages = {L687},
   title = {Welding of Transparent Materials Using Femtosecond Laser Pulses},
   volume = {44},
   url = {https://iopscience.iop.org/article/10.1143/JJAP.44.L687},
   year = {2005}
}

@article{Ozeki2008,
   author = {Yasuyuki Ozeki and Tomoyuki Inoue and Takayuki Tamaki and Hideaki Yamaguchi and Satoshi Onda and Wataru Watanabe and Tomokazu Sano and Shumpei Nishiuchi and Akio Hirose and Kazuyoshi Itoh},
   doi = {10.1143/APEX.1.082601},
   issn = {1882-0778},
   issue = {8},
   journal = {Applied Physics Express},
   month = {8},
   pages = {082601},
   title = {Direct Welding between Copper and Glass Substrates with Femtosecond Laser Pulses},
   volume = {1},
   url = {https://iopscience.iop.org/article/10.1143/APEX.1.082601},
   year = {2008}
}

@article{Yamaguchi2002,
   author = {K. Yamaguchi and K. Itagaki},
   doi = {10.1023/A:1020609517891},
   issn = {1388-6150},
   issue = {3},
   journal = {Journal of Thermal Analysis and Calorimetry},
   month = {7},
   pages = {1059-1066},
   title = {Measurement of high temperature heat content of silicon by drop calorimetry},
   volume = {69},
   url = {https://link.springer.com/10.1023/A:1020609517891},
   year = {2002}
}

@article{Chen2023,
   author = {Guoqing Chen and Xinyan Teng and Qianxing Yin and Binggang Zhang and Xuesong Leng},
   doi = {10.1007/s40194-022-01414-1},
   issn = {0043-2288},
   issue = {6},
   journal = {Welding in the World},
   month = {6},
   pages = {1479-1489},
   publisher = {Springer Science and Business Media Deutschland GmbH},
   title = {Effect of beam offset on the microstructure and properties of a dissimilar tungsten/kovar electron beam weld},
   volume = {67},
   url = {https://link.springer.com/10.1007/s40194-022-01414-1},
   year = {2023}
}

@misc{LamineriesMATTHEY,
   author = {Lamineries MATTHEY},
   title = {Physical properties of Kovar},
   url = {https://www.matthey.ch/fileadmin/user_upload/downloads/fichetechnique/EN/Kovar_v25E.pdf},
   year = {2025}
}

@article{Shanks1963,
   author = {H. R. Shanks and P. D. Maycock and P. H. Sidles and G. C. Danielson},
   doi = {10.1103/PhysRev.130.1743},
   issn = {0031-899X},
   issue = {5},
   journal = {Physical Review},
   month = {6},
   pages = {1743-1748},
   title = {Thermal Conductivity of Silicon from 300 to 1400°K},
   volume = {130},
   url = {https://link.aps.org/doi/10.1103/PhysRev.130.1743},
   year = {1963}
}

@article{Zanchetta1995,
   author = {Armando Zanchetta and Pierre Lefort and Emile Gabbay},
   doi = {10.1016/0955-2219(95)93944-X},
   issn = {09552219},
   issue = {3},
   journal = {Journal of the European Ceramic Society},
   month = {1},
   pages = {233-238},
   title = {Thermal expansion and adhesion of ceramic to metal sealings: Case of porcelain-kovar junctions},
   volume = {15},
   url = {https://linkinghub.elsevier.com/retrieve/pii/095522199593944X},
   year = {1995}
}

@article{Watanabe2004,
   author = {Hiromichi Watanabe and Naofumi Yamada and Masahiro Okaji},
   doi = {10.1023/B:IJOT.0000022336.83719.43},
   issn = {0195-928X},
   issue = {1},
   journal = {International Journal of Thermophysics},
   month = {1},
   pages = {221-236},
   title = {Linear Thermal Expansion Coefficient of Silicon from 293 to 1000 K},
   volume = {25},
   url = {https://link.springer.com/10.1023/B:IJOT.0000022336.83719.43},
   year = {2004}
}

@article{Carter2014,
   author = {Richard M. Carter and Jianyong Chen and Jonathan D. Shephard and Robert R. Thomson and Duncan P. Hand},
   doi = {10.1364/AO.53.004233},
   issn = {1559-128X},
   issue = {19},
   journal = {Applied Optics},
   month = {7},
   pages = {4233},
   publisher = {Optica Publishing Group},
   title = {Picosecond laser welding of similar and dissimilar materials},
   volume = {53},
   url = {https://opg.optica.org/abstract.cfm?URI=ao-53-19-4233},
   year = {2014}
}

@article{Shin2022,
   author = {Sungkwon Shin and Jun-Gyu Hur and Jong Kab Park and Doh-Hoon Kim},
   doi = {10.1364/OE.459377},
   issn = {1094-4087},
   issue = {11},
   journal = {Optics Express},
   month = {5},
   pages = {18018},
   pmid = {36221610},
   publisher = {Optica Publishing Group},
   title = {Polarization effects on ablation efficiency and microstructure symmetricity in femtosecond laser processing of materials—developing a pattern generation model for laser scanning},
   volume = {30},
   url = {https://opg.optica.org/abstract.cfm?URI=oe-30-11-18018},
   year = {2022}
}

@article{Mingareev2012,
   author = {Ilya Mingareev and Fabian Weirauch and Alexander Olowinsky and Lawrence Shah and Pankaj Kadwani and Martin Richardson},
   doi = {10.1016/j.optlastec.2012.03.020},
   issn = {00303992},
   issue = {7},
   journal = {Optics $\&$ Laser Technology},
   month = {10},
   pages = {2095-2099},
   title = {Welding of polymers using a 2 $\mu$m thulium fiber laser},
   volume = {44},
   url = {https://linkinghub.elsevier.com/retrieve/pii/S0030399212001302},
   year = {2012}
}

@article{Zhang2015,
   author = {Guodong Zhang and Guanghua Cheng},
   doi = {10.1364/AO.54.008957},
   issn = {0003-6935},
   issue = {30},
   journal = {Applied Optics},
   month = {10},
   pages = {8957},
   pmid = {26560385},
   title = {Direct welding of glass and metal by 1 kHz femtosecond laser pulses},
   volume = {54},
   url = {https://opg.optica.org/abstract.cfm?URI=ao-54-30-8957},
   year = {2015}
}

@article{Richter2016,
   author = {Sören Richter and Felix Zimmermann and Andreas Tünnermann and Stefan Nolte},
   doi = {10.1016/j.optlastec.2016.03.022},
   issn = {00303992},
   journal = {Optics $\&$ Laser Technology},
   month = {9},
   pages = {59-66},
   publisher = {Elsevier Ltd},
   title = {[INVITED] Laser welding of glasses at high repetition rates – Fundamentals and prospects},
   volume = {83},
   url = {https://linkinghub.elsevier.com/retrieve/pii/S0030399215305958},
   year = {2016}
}

@article{Penilla2019,
   author = {E. H. Penilla and L. F. Devia-Cruz and A. T. Wieg and P. Martinez-Torres and N. Cuando-Espitia and P. Sellappan and Y. Kodera and G. Aguilar and J. E. Garay},
   doi = {10.1126/science.aaw6699},
   issn = {0036-8075},
   issue = {6455},
   journal = {Science},
   month = {8},
   pages = {803-808},
   title = {Ultrafast laser welding of ceramics},
   volume = {365},
   url = {https://www.science.org/doi/10.1126/science.aaw6699},
   year = {2019}
}

@article{Cvecek2019,
   author = {Kristian Cvecek and Sarah Dehmel and Isamu Miyamoto and Michael Schmidt},
   doi = {10.1088/2631-7990/ab55f6},
   issn = {2631-8644},
   issue = {4},
   journal = {International Journal of Extreme Manufacturing},
   month = {12},
   pages = {042001},
   publisher = {IOP Publishing Ltd},
   title = {A review on glass welding by ultra-short laser pulses},
   volume = {1},
   url = {https://iopscience.iop.org/article/10.1088/2631-7990/ab55f6},
   year = {2019}
}

@article{Jiang2025,
   author = {Yu-Guo Jiang and Jia-Fan Kuo and Chung-Wei Cheng and An-Chen Lee and Yasuhiro Okamoto},
   doi = {10.1016/j.optlastec.2025.112489},
   issn = {00303992},
   journal = {Optics $\&$ Laser Technology},
   month = {6},
   pages = {112489},
   publisher = {Elsevier Ltd},
   title = {Direct welding of silicon carbide and fused silica using femtosecond lasers: Effects of repetition rate and focal depth},
   volume = {184},
   url = {https://linkinghub.elsevier.com/retrieve/pii/S0030399225000775},
   year = {2025}
}

@article{Zhang2018,
   author = {Guodong Zhang and Razvan Stoian and Wei Zhao and Guanghua Cheng},
   doi = {10.1364/OE.26.000917},
   issn = {1094-4087},
   issue = {2},
   journal = {Optics Express},
   month = {1},
   pages = {917},
   publisher = {Optica Publishing Group},
   title = {Femtosecond laser Bessel beam welding of transparent to non-transparent materials with large focal-position tolerant zone},
   volume = {26},
   url = {https://opg.optica.org/abstract.cfm?URI=oe-26-2-917},
   year = {2018}
}

@article{Ji2025,
   author = {Changhao Ji and Cheng Yang and Chengyun Wang and Ziyue Yu and Jindou Wu and Lili Yuan and Yu Long},
   doi = {10.1016/j.optlastec.2025.113570},
   issn = {00303992},
   journal = {Optics $\&$ Laser Technology},
   month = {12},
   pages = {113570},
   publisher = {Elsevier Ltd},
   title = {A review of the progress and challenges in ultrafast laser micro-welding of brittle optical materials},
   volume = {192},
   url = {https://linkinghub.elsevier.com/retrieve/pii/S0030399225011612},
   year = {2025}
}

@article{Chanal2017,
   author = {Margaux Chanal and Vladimir Yu. Fedorov and Maxime Chambonneau and Raphaël Clady and Stelios Tzortzakis and David Grojo},
   doi = {10.1038/s41467-017-00907-8},
   issn = {2041-1723},
   issue = {1},
   journal = {Nature Communications},
   month = {10},
   pages = {773},
   pmid = {28974678},
   publisher = {Nature Publishing Group},
   title = {Crossing the threshold of ultrafast laser writing in bulk silicon},
   volume = {8},
   url = {https://www.nature.com/articles/s41467-017-00907-8},
   year = {2017}
}

@article{Wang2025,
   author = {Andong Wang and Amlan Das and Vladimir Yu Fedorov and Pol Sopeña and Stelios Tzortzakis and David Grojo},
   doi = {10.1038/s41467-025-61983-9},
   issn = {2041-1723},
   issue = {1},
   journal = {Nature Communications},
   month = {7},
   pages = {6733},
   pmid = {40695822},
   publisher = {Nature Research},
   title = {In-chip critical plasma seeds for laser writing of reconfigurable silicon photonics systems},
   volume = {16},
   url = {https://www.nature.com/articles/s41467-025-61983-9},
   year = {2025}
}

@article{Sopena2022,
   author = {Pol Sopeña and Andong Wang and Alexandros Mouskeftaras and David Grojo},
   doi = {10.1002/lpor.202200208},
   issn = {1863-8880},
   issue = {11},
   journal = {Laser $\&$ Photonics Reviews},
   month = {11},
   publisher = {John Wiley and Sons Inc},
   title = {Transmission Laser Welding of Similar and Dissimilar Semiconductor Materials},
   volume = {16},
   url = {https://onlinelibrary.wiley.com/doi/10.1002/lpor.202200208},
   year = {2022}
}

@article{Sari2008,
   author = {F. Sari and W.-M. Hoffmann and E. Haberstroh and R. Poprawe},
   doi = {10.1007/s00542-008-0675-3},
   issn = {0946-7076},
   issue = {12},
   journal = {Microsystem Technologies},
   month = {11},
   pages = {1879-1886},
   title = {Applications of laser transmission processes for the joining of plastics, silicon and glass micro parts},
   volume = {14},
   url = {http://link.springer.com/10.1007/s00542-008-0675-3},
   year = {2008}
}

@article{Li2021,
   author = {Qingfeng Li and Maxime Chambonneau and Markus Blothe and Herbert Gross and Stefan Nolte},
   doi = {10.1364/AO.421945},
   issn = {1559-128X},
   issue = {13},
   journal = {Applied Optics},
   month = {5},
   pages = {3954},
   pmid = {33983334},
   publisher = {Optica Publishing Group},
   title = {Flexible, fast, and benchmarked vectorial model for focused laser beams},
   volume = {60},
   url = {https://opg.optica.org/abstract.cfm?URI=ao-60-13-3954},
   year = {2021}
}

@misc{Li2021a,
   author = {Qingfeng Li},
   title = {InFocus},
   url = {https://github.com/QF06/InFocus},
   year = {2021}
}

@misc{standarddamage,
   author = {International Organization for Standardization},
   edition = {1},
   issue = {ISO 21254-2:2011},
   month = {7},
   note = {Published 2011-07; last reviewed and confirmed in 2021},
   title = {ISO 21254-2:2011 Lasers and laser-related equipment – Test methods for laser-induced damage threshold – Part 2: Threshold determination},
   url = {https://www.iso.org/standard/43002.html},
   year = {2011}
}

@misc{standardleakrate,
   author = {DIN Deutsches Institut für Normung e. V.},
   doi = {10.31030/3577042},
   issue = {DIN EN 1779:2024-12},
   month = {12},
   note = {Draft standard; date of issue 2024-11-15; see Annex B, Section B.2.2},
   publisher = {DIN Media GmbH},
   title = {DIN EN 1779:2024-12 Non-destructive testing – Leak testing – Criteria for method and technique selection; German and English version prEN 1779:2024},
   url = {https://www.dinmedia.de/en/draft-standard/din-en-1779/384653977},
   year = {2024}
}

@article{Su2025,
   author = {Xiaohui Su and Si Wu and Jian Shao and Xinke Xu and Zijing Yang and Yi Liu and Yaqing Qiao and Qiaodan Chen and Leimin Deng},
   doi = {10.1364/OE.559168},
   issn = {1094-4087},
   issue = {11},
   journal = {Optics Express},
   month = {6},
   pages = {22469},
   pmid = {40515235},
   publisher = {Optica Publishing Group},
   title = {Flexible modulation of molten areas by fs-laser discrete scanning glass welding},
   volume = {33},
   url = {https://opg.optica.org/abstract.cfm?URI=oe-33-11-22469},
   year = {2025}
}

@article{Wang2025b,
   author = {Yanbin Wang and Mingzhi Yu and Yao Chen and Yintao Ma and Xiangguang Han and Yong Xia and Ju Guo and Ping Yang and Qijing Lin and Shujiang Ding and Libo Zhao},
   doi = {10.1038/s41378-025-00976-6},
   issn = {2055-7434},
   issue = {1},
   journal = {Microsystems $\&$ Nanoengineering},
   month = {8},
   pages = {153},
   publisher = {Springer Nature},
   title = {The fabrication of MEMS alkali metal vapor cells based on ultrafast laser welding for single beam magnetometer},
   volume = {11},
   url = {https://www.nature.com/articles/s41378-025-00976-6},
   year = {2025}
}

@article{Gstalter2019,
   author = {Marion Gstalter and Grégoire Chabrol and Armel Bahouka and Kokou-Dodzi Dorkenoo and Jean-Luc Rehspringer and Sylvain Lecler},
   doi = {10.1364/AO.58.008858},
   issn = {1559-128X},
   issue = {32},
   journal = {Applied Optics},
   month = {11},
   pages = {8858},
   pmid = {31873671},
   publisher = {Optica Publishing Group},
   title = {Long focal length high repetition rate femtosecond laser glass welding},
   volume = {58},
   url = {https://opg.optica.org/abstract.cfm?URI=ao-58-32-8858},
   year = {2019}
}

@article{Hecker2020a,
   author = {Sebastian Hecker and Markus Blothe and Daniel Grossmann and Thomas Graf},
   doi = {10.1364/AO.392702},
   issn = {1559-128X},
   issue = {22},
   journal = {Applied Optics},
   month = {8},
   pages = {6452},
   pmid = {32749342},
   publisher = {Optica Publishing Group},
   title = {Process regimes during welding of glass by femtosecond laser pulse bursts},
   volume = {59},
   url = {https://opg.optica.org/abstract.cfm?URI=ao-59-22-6452},
   year = {2020}
}

@article{Gstalter2017,
   author = {M. Gstalter and G. Chabrol and A. Bahouka and L. Serreau and J-L. Heitz and G. Taupier and K-D. Dorkenoo and J-L. Rehspringer and S. Lecler},
   doi = {10.1007/s00339-017-1324-4},
   issn = {0947-8396},
   issue = {11},
   journal = {Applied Physics A},
   month = {11},
   pages = {714},
   publisher = {Springer Verlag},
   title = {Stress-induced birefringence control in femtosecond laser glass welding},
   volume = {123},
   url = {http://link.springer.com/10.1007/s00339-017-1324-4},
   year = {2017}
}

@article{Hecker2020b,
   author = {Sebastian Hecker and Markus Blothe and Thomas Graf},
   doi = {10.1364/AO.411667},
   issn = {1559-128X},
   issue = {36},
   journal = {Applied Optics},
   month = {12},
   pages = {11382},
   pmid = {33362063},
   publisher = {Optica Publishing Group},
   title = {Reproducible process regimes during glass welding by bursts of subpicosecond laser pulses},
   volume = {59},
   url = {https://opg.optica.org/abstract.cfm?URI=ao-59-36-11382},
   year = {2020}
}

@article{Xie2026,
   author = {Qiong Xie and Niladri Ganguly and Pol Sopeña and David Grojo},
   doi = {10.1063/5.0334492},
   issn = {0003-6951},
   issue = {22},
   journal = {Applied Physics Letters},
   month = {6},
   publisher = {American Institute of Physics},
   title = {Persistence of self-limited conditions in bulk silicon induced by mid-infrared femtosecond laser pulses},
   volume = {128},
   url = {https://pubs.aip.org/apl/article/128/22/221105/3393545/Persistence-of-self-limited-conditions-in-bulk},
   year = {2026}
}

@article{Ganguly2024,
   author = {Niladri Ganguly and Pol Sopeña and David Grojo},
   doi = {10.37188/lam.2024.022},
   issn = {2831-4093},
   issue = {3},
   journal = {Light: Advanced Manufacturing},
   pages = {1},
   publisher = {Light Publishing Group},
   title = {Ultra-high-aspect-ratio structures through silicon using infrared laser pulses focused with axicon-lens doublets},
   volume = {5},
   url = {https://www.light-am.com/article/doi/10.37188/lam.2024.022},
   year = {2024}
}

@article{Jia2025,
   author = {Xianshi Jia and Jinlin Luo and Kai Li and Cong Wang and Zhou Li and Mengmeng Wang and Zhengyi Jiang and Vadim P Veiko and Ji’an Duan},
   doi = {10.1088/2631-7990/ada7a7},
   issn = {2631-8644},
   issue = {3},
   journal = {International Journal of Extreme Manufacturing},
   month = {6},
   pages = {032001},
   publisher = {Institute of Physics},
   title = {Ultrafast laser welding of transparent materials: from principles to applications},
   volume = {7},
   url = {https://iopscience.iop.org/article/10.1088/2631-7990/ada7a7},
   year = {2025}
}

@article{Bleiker2017,
   author = {S.J. Bleiker and V. Dubois and S. Schröder and G. Stemme and F. Niklaus},
   doi = {10.1016/j.sna.2017.04.018},
   issn = {09244247},
   journal = {Sensors and Actuators A: Physical},
   month = {6},
   pages = {16-23},
   publisher = {Elsevier B.V.},
   title = {Adhesive wafer bonding with ultra-thin intermediate polymer layers},
   volume = {260},
   url = {https://linkinghub.elsevier.com/retrieve/pii/S0924424717301292},
   year = {2017}
}

@inproceedings{Reuter2005,
   author = {Danny S. Reuter and Andreas Bertz and Gunther Schwenzer and Thomas Gessner},
   doi = {10.1117/12.581829},
   editor = {Jung-Chih Chiao and David N. Jamieson and Lorenzo Faraone and Andrew S. Dzurak},
   issn = {0277786X},
   booktitle = {Proceedings of SPIE},
   month = {2},
   pages = {163},
   publisher = {SPIE},
   title = {Selective adhesive bonding with SU-8 for zero-level-packaging},
   volume = {5650},
   url = {http://proceedings.spiedigitallibrary.org/proceeding.aspx?doi=10.1117/12.581829},
   year = {2005}
}

@article{Abouie2012,
   author = {Maryam Abouie and Qi Liu and Douglas G. Ivey},
   doi = {10.1016/j.mseb.2012.09.005},
   issn = {09215107},
   issue = {20},
   journal = {Materials Science and Engineering: B},
   month = {12},
   pages = {1748-1758},
   publisher = {Elsevier Ltd},
   title = {Eutectic and solid-state wafer bonding of silicon with gold},
   volume = {177},
   url = {https://linkinghub.elsevier.com/retrieve/pii/S0921510712004692},
   year = {2012}
}

@article{Jing2010,
   author = {Errong Jing and Bin Xiong and Yuelin Wang},
   doi = {10.1109/TEPM.2009.2035307},
   issn = {1521-334X},
   issue = {1},
   journal = {IEEE Transactions on Electronics Packaging Manufacturing},
   month = {1},
   pages = {31-37},
   title = {The Bond Strength of Au/Si Eutectic Bonding Studied by IR Microscope},
   volume = {33},
   url = {http://ieeexplore.ieee.org/document/5371921/},
   year = {2010}
}

@article{Ding2022,
   author = {Zhijie Ding and Haitao Xue and Weibing Guo and Cuixin Chen and Yang Jia},
   doi = {10.1016/j.mtcomm.2022.104451},
   issn = {23524928},
   journal = {Materials Today Communications},
   month = {12},
   pages = {104451},
   publisher = {Elsevier Ltd},
   title = {Microstructure and shear strength on ultrasonic-assisted soldered the Si/Cu joint using Sn-3.5Ag solder},
   volume = {33},
   url = {https://linkinghub.elsevier.com/retrieve/pii/S2352492822012922},
   year = {2022}
}

@article{Kolenak2021,
   author = {Roman Kolenak and Igor Kostolny and Jaromir Drapala and Paulina Babincova and Peter Gogola},
   doi = {10.3390/met11040624},
   issn = {2075-4701},
   issue = {4},
   journal = {Metals},
   month = {4},
   pages = {624},
   publisher = {MDPI AG},
   title = {Characterization of Soldering Alloy Type Bi-Ag-Ti and the Study of Ultrasonic Soldering of Silicon and Copper},
   volume = {11},
   url = {https://www.mdpi.com/2075-4701/11/4/624},
   year = {2021}
}

@article{Kolenak2016,
   author = {Roman Kolenak and Igor Kostolný and Martin Sahul},
   doi = {10.1108/SSMT-11-2015-0040},
   issn = {0954-0911},
   issue = {3},
   journal = {Soldering $\&$ Surface Mount Technology},
   month = {6},
   pages = {149-158},
   publisher = {Emerald Group Publishing Ltd.},
   title = {Direct bonding of silicon with solders type Sn-Ag-Ti},
   volume = {28},
   url = {http://www.emerald.com/ssmt/article/28/3/149-158/365430},
   year = {2016}
}

@article{Li2023,
   author = {Wenzhao Li and Zhijie Ding and Haitao Xue and Weibing Guo and Cuixin Chen and Yang Jia and Zheng Wan},
   doi = {10.1016/j.matchar.2023.112833},
   issn = {10445803},
   journal = {Materials Characterization},
   month = {5},
   pages = {112833},
   publisher = {Elsevier Inc.},
   title = {Interfacial bonding mechanisms in ultrasonic-assisted soldered Si/Cu joint using Sn-3.5Ag-4Al solder},
   volume = {199},
   url = {https://linkinghub.elsevier.com/retrieve/pii/S1044580323001912},
   year = {2023}
}

@article{Xue2024,
   author = {Haitao Xue and Zheng Wan and Zhijie Ding and Weibing Guo and Yang Jia and Cuixin Chen and Fuxing Yin and Wenzhao Li and Wenjie Mu},
   doi = {10.1016/j.msea.2023.146033},
   issn = {09215093},
   journal = {Materials Science and Engineering: A},
   month = {2},
   pages = {146033},
   publisher = {Elsevier Ltd},
   title = {Ultrasonic-assisted soldering of Si and Cu joint with SnAgTi solder in air},
   volume = {892},
   url = {https://linkinghub.elsevier.com/retrieve/pii/S0921509323014570},
   year = {2024}
}

@article{Keyvaninia2013,
   author = {S. Keyvaninia and M. Muneeb and S. Stanković and P. J. Van Veldhoven and D. Van Thourhout and G. Roelkens},
   doi = {10.1364/OME.3.000035},
   issn = {2159-3930},
   issue = {1},
   journal = {Optical Materials Express},
   month = {1},
   pages = {35},
   publisher = {IEEE},
   title = {Ultra-thin DVS-BCB adhesive bonding of III-V wafers, dies and multiple dies to a patterned silicon-on-insulator substrate},
   volume = {3},
   url = {https://opg.optica.org/ome/abstract.cfm?uri=ome-3-1-35},
   year = {2013}
}

@article{Stankovic2011,
   author = {S. Stankovi\'c and R. Jones and J. Heck and M. Sysak and D. Van Thourhout and G. Roelkens},
   doi = {10.1149/1.3592267},
   issn = {10990062},
   issue = {8},
   journal = {Electrochemical and Solid-State Letters},
   pages = {H326},
   title = {Die-to-Die Adhesive Bonding Procedure for Evanescently-Coupled Photonic Devices},
   volume = {14},
   url = {https://iopscience.iop.org/article/10.1149/1.3592267},
   year = {2011}
}

@article{Zavedeev2016,
   author = {E V Zavedeev and V V Kononenko and V I Konov},
   doi = {10.1088/1054-660X/26/1/016101},
   issn = {1054-660X},
   issue = {1},
   journal = {Laser Physics},
   month = {1},
   pages = {016101},
   publisher = {Institute of Physics Publishing},
   title = {Delocalization of femtosecond laser radiation in crystalline Si in the mid-IR range},
   volume = {26},
   url = {https://iopscience.iop.org/article/10.1088/1054-660X/26/1/016101},
   year = {2016}
}

@article{Mareev2020,
   author = {E I Mareev and K V Lvov and B V Rumiantsev and E A Migal and I D Novikov and S Yu Stremoukhov and F V Potemkin},
   doi = {10.1088/1612-202X/ab5d23},
   issn = {1612-2011},
   issue = {1},
   journal = {Laser Physics Letters},
   month = {1},
   pages = {015402},
   publisher = {Institute of Physics Publishing},
   title = {Effect of pulse duration on the energy delivery under nonlinear propagation of tightly focused Cr:forsterite laser radiation in bulk silicon},
   volume = {17},
   url = {https://iopscience.iop.org/article/10.1088/1612-202X/ab5d23},
   year = {2020}
}

@article{vanAbeelen2026,
   author = {Tara van Abeelen and Lucas Groult and Richard M. Carter and M. J. Daniel Esser and Duncan P. Hand},
   doi = {10.1364/OME.606379},
   issn = {2159-3930},
   issue = {8},
   journal = {Optical Materials Express},
   month = {8},
   pages = {2570},
   title = {2 $\mu$m wavelength nanosecond laser welding of silicon},
   volume = {16},
   url = {https://opg.optica.org/abstract.cfm?URI=ome-16-8-2570},
   year = {2026}
}

@article{Capodacqua2023,
   author = {Filippo Maria Conte Capodacqua and Annalisa Volpe and Caterina Gaudiuso and Antonio Ancona},
   doi = {10.1038/s41598-023-31969-y},
   issn = {2045-2322},
   issue = {1},
   journal = {Scientific Reports},
   month = {3},
   pages = {5062},
   pmid = {36977765},
   publisher = {Nature Research},
   title = {Bonding of PMMA to silicon by femtosecond laser pulses},
   volume = {13},
   url = {https://www.nature.com/articles/s41598-023-31969-y},
   year = {2023}
}
}

\end{document}